\documentclass[%
 aip,
 amsmath,amssymb,
 reprint,%
]{revtex4-1}

\usepackage[per-mode=symbol,separate-uncertainty]{siunitx}
\usepackage[english]{babel}

\usepackage{graphicx}% Include figure files
\usepackage{dcolumn}% Align table columns on decimal point
\usepackage{bm}% bold math
\usepackage[protrusion=true, expansion=true]{microtype} % improved typesetting in pdflatex
\usepackage[
	pdffitwindow=true,
	colorlinks=true,
	frenchlinks=false,
    linkcolor=blue,
    anchorcolor=blue,
    citecolor=blue,
    filecolor=blue,
    urlcolor=blue,
    bookmarks=true,
    bookmarksopen=true,
	bookmarksnumbered=true, 
    bookmarksopenlevel=1,
    plainpages=false,
	pdfpagelayout=SinglePage,
    pdfpagelabels=true,
	breaklinks
]{hyperref}

\usepackage[utf8]{inputenc}
\usepackage[T1]{fontenc}
\usepackage{mathptmx}
\usepackage{etoolbox}
\usepackage{subfigure}

\newcommand\bra[1]{{\langle{#1}|}} %JG remove space between \ket and \bra
\makeatletter
\newcommand\ket[1]{%
  \@ifnextchar\bra{\k@t{#1}\!}{\k@t{#1}}%
}
\newcommand\k@t[1]{{|{#1}\rangle}}
\makeatother

\makeatletter
\def\@email#1#2{%
 \endgroup
 \patchcmd{\titleblock@produce}
  {\frontmatter@RRAPformat}
  {\frontmatter@RRAPformat{\produce@RRAP{*#1\href{mailto:#2}{#2}}}\frontmatter@RRAPformat}
  {}{}
}%
\makeatother
\begin{document}

\title[Charge dynamics in quantum-circuit refrigeration]{Charge dynamics in quantum-circuit refrigeration: \\ thermalization and microwave gain}
 
\author{Hao Hsu}
\affiliation{JARA Institute for Quantum Information (PGI-11), Forschungszentrum J{\"u}lich, \\ 52425 J{\"u}lich, Germany}
\author{Matti Silveri}
\affiliation{Nano and Molecular Systems Research Unit, University of Oulu, \\ P.O.~Box 3000, FI-90014 Oulu, Finland}
\author{Vasilii Sevriuk}
\affiliation{QCD Labs, QTF Centre of Excellence, Department of Applied Physics, Aalto University, \\ P.O.~Box 13500, FI-00076 Aalto, Finland}
\affiliation{IQM, Keilaranta 19, 02150 Espoo, Finland}
\author{Mikko M\"{o}tt\"{o}nen}
\affiliation{QCD Labs, QTF Centre of Excellence, Department of Applied Physics, Aalto University, \\ P.O.~Box 13500, FI-00076 Aalto, Finland}
\affiliation{QTF Centre of Excellence, VTT Technical Research Centre of Finland Ltd,\\ P.O. Box 1000, FI-02044 VTT, Finland}
\author{Gianluigi Catelani}
\affiliation{JARA Institute for Quantum Information (PGI-11), Forschungszentrum J{\"u}lich, \\ 52425 J{\"u}lich, Germany}

\date{\today}% It is always \today, today,
             %  but any date may be explicitly specified

\begin{abstract}
Previous studies of photon-assisted tunneling through normal-metal–insulator–superconductor
junctions have exhibited potential for providing a convenient tool to control the dissipation of quantum-electric
circuits in-situ. However, the current literature on such a quantum-circuit refrigerator (QCR) does not present a detailed
description for the charge dynamics of the tunneling processes or the phase coherence of the open quantum system. Here we derive a master equation describing both quantum-electric and charge degrees of freedom,
and discover that typical experimental parameters of low temperature and yet lower charging energy yield a separation
of time scales for the charge and quantum dynamics. Consequently, the minor effect of the different charge states can
be taken into account by averaging over the charge distribution. We also consider applying an ac
voltage to the tunnel junction, which enables control of the decay rate of a superconducting qubit over four orders of magnitude by changing the drive amplitude; we find an order-of-magnitude drop in the qubit
excitation in 40~ns and a residual reset infidelity below $10^{-4}$. Furthermore, for the normal island we consider the case of charging
energy and single-particle level spacing large compared to the superconducting gap, i.e., a quantum dot. Although the decay rates
arising from such a dot QCR appear low for use in qubit reset, the device can
provide effective negative damping (gain) to the coupled microwave resonator. The Fano factor
of such a millikelvin microwave source may be smaller than unity, with the latter value being reached close to the maximum attainable power.

\end{abstract}

\maketitle

\section{Introduction}
\label{sec:intro}

The quantum-circuit refrigerator is a hybrid quantum device made of a normal-metal island tunnel-coupled to two voltage-biased superconducting leads.\cite{tan_2017, Silveri_2017} The device to be refrigerated is capacitively coupled to the island. When a tunneling event occurs between the island and a superconducting lead, it can be supplemented by a photon transfer between the coupled device and the tunneling electron. Depending on the voltage bias, the quantum-circuit refrigerator (QCR) has either a cooling or a heating effect on the coupled device, which can be considered to originate from the energy-filtering effect in the tunneling process promoted by the gaped superconductor density of states at the superconductor\---insulator\---normal-metal junctions. 

In the realm of superconducting quantum circuits, the QCR is a powerful tool to control the effective environment of the coupled quantum-electric device, with applications in quantum heat transfer,\cite{partanen2019} fundamental physics explorations,\cite{Silveri_2019} and rapid state reset for quantum information applications.\cite{partanen2018, sevriuk_2019, hsu_2020} The QCR is a part of the recent inspiring theoretical approaches to open quantum systems with applications to quantum devices\cite{clerk2013,murch2019,kamal2020,kamal2021} and experiments related to the quantum reservoir engineering, promoting its usefulness for control and simulation of quantum-information systems.\cite{mostame_2012,jones_2013, tuorila_2017}

Previous theoretical analyses focused on the QCR-induced transition rates in a harmonic resonator\cite{Silveri_2017} or a weakly anharmonic qubit.\cite{hsu_2020} The underlying assumption was that the charging energy of the normal-metal island is sufficiently small so that possible effects by normal-metal charge dynamics can be safely neglected. In this approach, the QCR-induced transition rates can be interpreted to originate from a bath, the temperature and coupling strength of which are voltage-bias tunable, with excellent agreement in comparison to experimental results.\cite{Silveri_2017, Silveri_2019, sevriuk_2019, hyyppa_2019} In this work, we go beyond such an approach and study in more detail the charge dynamics of the QCR by deriving an extended master equation which accounts for both the coupled quantum device and the QCR normal-metal island. This enables us to understand the effect of the charge dynamics on quantum-circuit refrigeration. 

We focus on two experimentally relevant cases. First, we derive the extended master equation for a highly anharmonic superconducting qubit. As a result we obtain an explicit expression for the time scales of charge relaxation in the normal-metal island. With typical experimental parameters, i.e., for the charging energy of the normal-metal island much smaller than the thermal energy set by electron temperature and for the QCR biased in the cooling regime, the charge relaxation rate is at least an order of magnitude slower than the QCR-induced decay rates in the qubit. This implies that during QCR operation the normal-island has essentially no charge dynamics and its contributions to the transition rates can be safely averaged, in agreement with the assumptions of Refs.~\onlinecite{Silveri_2017, hsu_2020}. 

In the second part, we treat the other extreme limit of the charge dynamics, namely, the case in which the QCR normal-metal island is reduced in size such that the charging energy and the single-particle level spacing are larger than all other relevant energy scales in the system. Here, the island can host a single electron at most. In other words, we consider a single-level quantum dot coupled to two superconducting leads. If such a dot QCR is capacitively coupled to a resonator, there are operational regimes, in which a non-classical state of light can be induced in the resonator
and spontaneous microwave generation is possible, a situation resembling that of single-atom lasers.\cite{mu_1992} Here, the microwave generation originates from the peaked density of states of the superconductor, enabling bias-voltage controlled operation regimes where the QCR-induced excitation dominates over decay, resulting in so-called negative damping.  

Focusing on low-temperature solid-state devices, we can roughly classify cryogenic microwave sources by the type of active environment: they can be based on superconducting qubits,\cite{you_2007,astafiev_2007,Ashhab_2009,hauss_2008,astafiev_2021} single-electron transistors,\cite{set_2007,marthaler_2011} or quantum dots.\cite{stockklauser_2015,Jin_2013,liu_2015,Liu_2017threshold,Liu_2017phase,xu_2013} Most of the works related to quantum dots are based on the electron tunneling between two dots whose levels are detuned by applied gate voltages. The energy of the detuning is equal to the energy of the generated photons. Such design is limited by the low tunneling rates and calls for advanced control and fabrication techniques. In this paper, we discuss microwave generation based on a single quantum dot (QD) coupled to two superconducting leads through tunnel barriers, usually thin insulating layers. Such structures are particularly interesting due to the counter-intuitive exploitation of a non-monotonic quasiparticle density of states for amplification.

This idea was previously discussed in Ref.~\onlinecite{bruhat_2016}, where a co-planar-waveguide resonator is coupled to the QD which is attached to superconducting and normal-metal leads. The authors also reported negative damping in this system. However, no estimation of the damping rate was given, and no detailed description of the distinction between photon generation and negative damping was provided. Such distinction is important since it is possible to observe photon generation not related to the negative damping rate, i.e., heating,  as we reported in Ref.~\onlinecite{Masuda_2016}. A similar system to that studied in this work has been also experimentally realized in a single-electron turnstile study.\cite{Denis_2016} The superconducting leads were formed from a superconducting wire by electrochemical migration and a metallic nanoparticle deposited from a solution was used as the QD. The system was equipped with a bottom gate to tune the QD potential. More details about the fabrication can be found in Refs.~\onlinecite{zanten2015fab_iv, park_1999fab}. Realizations with nanowire-based quantum dots are reported in Ref.~\onlinecite{doh}; however, these systems had no resonator coupled to the~QD.

The paper is organized as follows. In Sec.~\ref{sec:QCR}, we introduce a model for a QCR coupled to a qubit and present the corresponding master equation governing the dynamics of the qubit and the charge states of the QCR. The master equation is solved in experimentally relevant regimes where the charging energy of the QCR is small compared with its electron temperature. We also study the case of a classical alternating-current (ac) voltage control of the QCR as an alternative to the pure dc bias (see also Ref.~\onlinecite{viitanen}).
In Sec.~\ref{sec:dotQCR}, we modify our model to treat the case of the dot QCR coupled to a resonator, analyze its suitability for cooling a resonator, and study in detail the regime of microwave signal generation.
We summarize our findings in Sec.~\ref{sec:conclusions}. Appendices~\ref{app:2nd}--\ref{app:totdec} contain mathematical details relating to Sec.~\ref{sec:QCR}.

\section{QCR coupled to a two-level system}
\label{sec:QCR}

\begin{figure}[tb]
\centering
\subfigure{
\includegraphics[width=3.3in,height=2.8in]{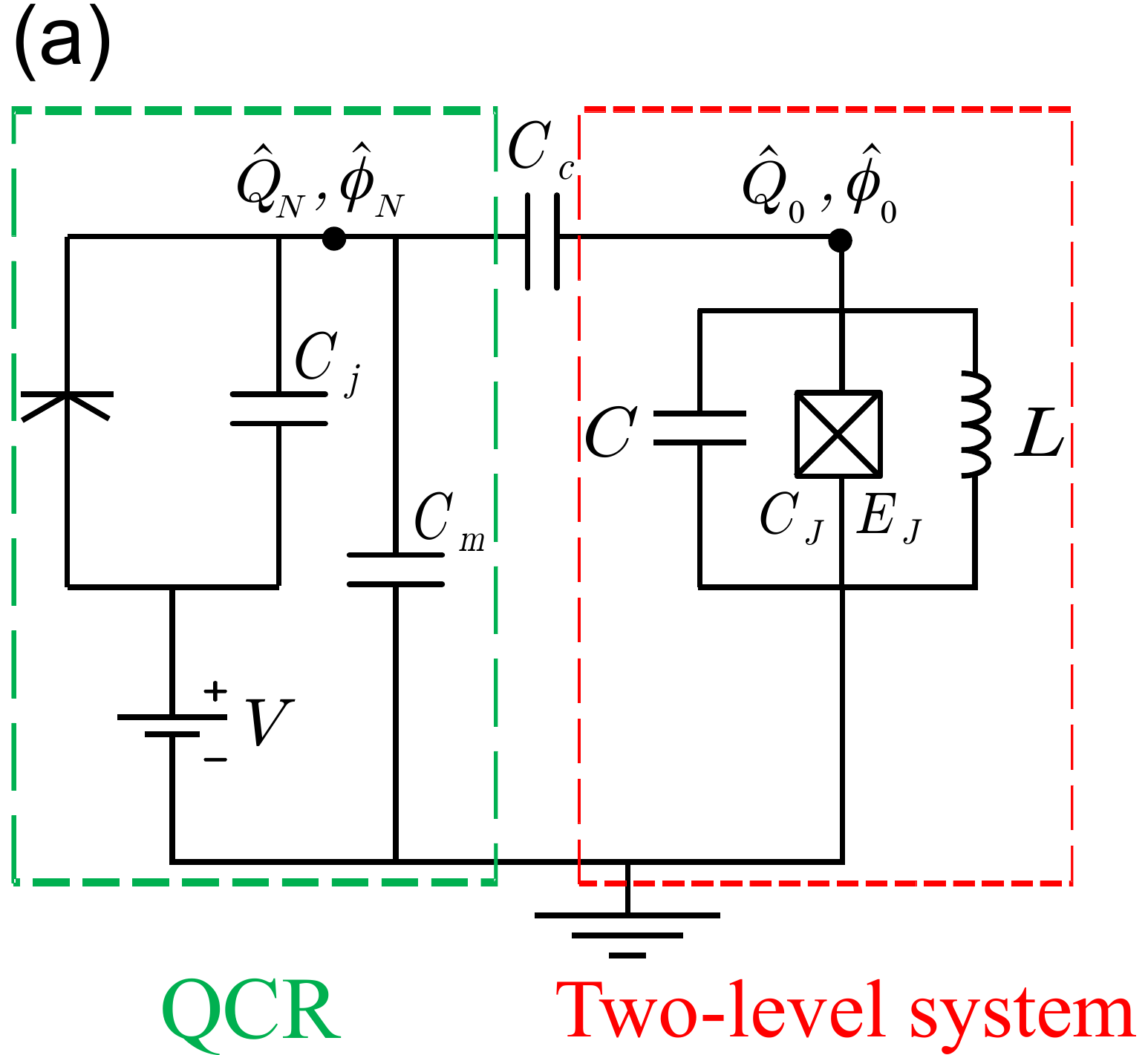}
}
\subfigure{
\includegraphics[width=3.3in]{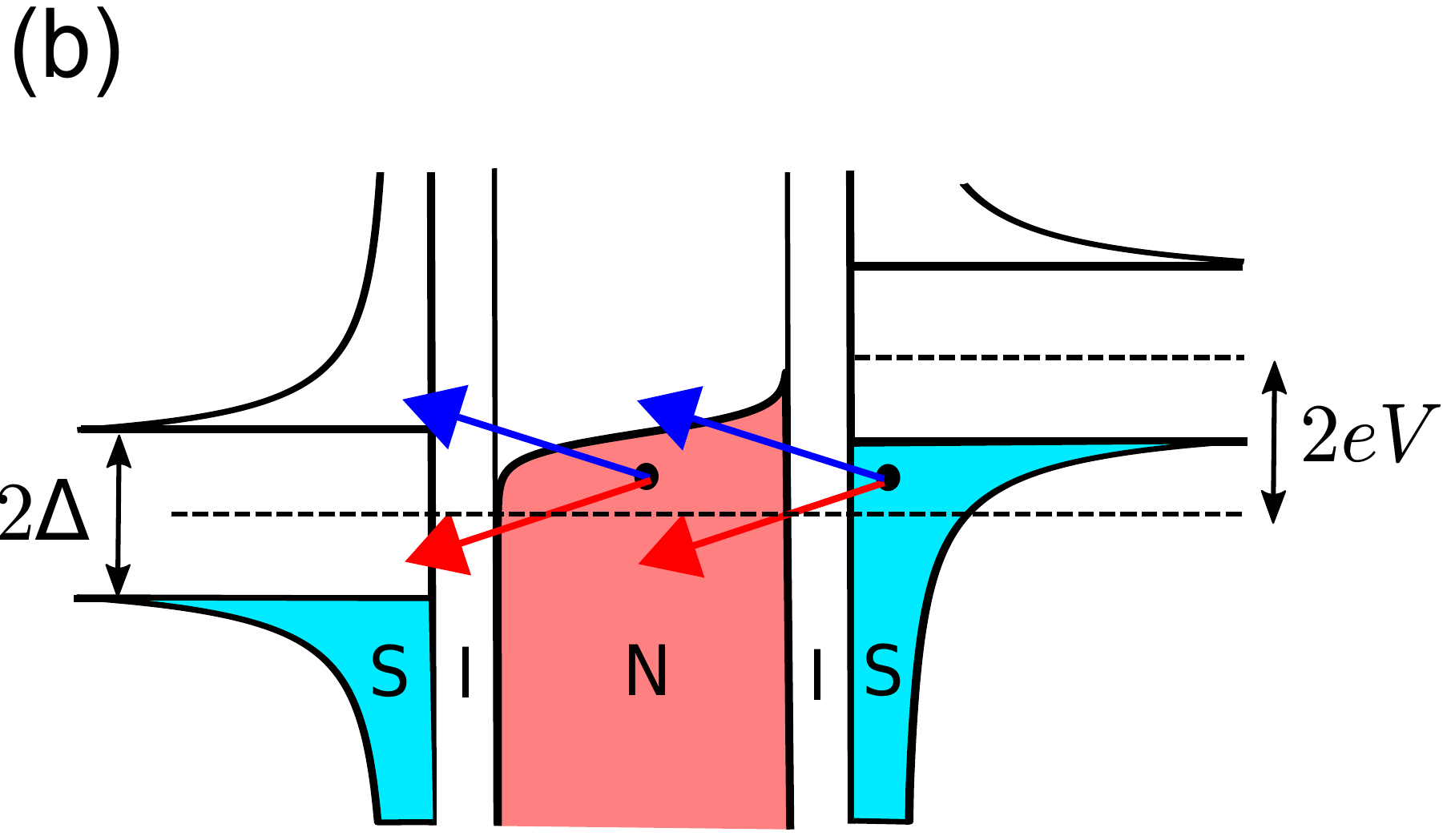} 
}
\caption{(a)~Effective circuit diagram for a quantum-circuit refrigerator (QCR) capacitively coupled to a two-level system consisting of a capacitor, a Josephson junction through the capacitance $C_c$. (b)~Schematic diagram of the density of states of the voltage biased superconductor--insulator--normal-metal--insulator--superconductor~(SINIS) junction. Photon-assisted electron tunnelings are depicted by colorful arrows, such as a blue arrow depicts a tunneling event which the SINIS junction absorbs a photon from the coupled two-level system. We refer to the bias-voltage-controlled and capacitively coupled SINIS junction as the quantum-circuit refrigerator~(QCR).}
\label{fig:QCR_2level}
\end{figure}

We consider a system in which a QCR is capacitively coupled to a highly anharmonic superconducting qubit, as schematically depicted in Fig.~\ref{fig:QCR_2level}.
The Hamiltonian modeling the system consists of four terms,%~\cite{hsu_2020}
\begin{equation}\label{eq:Htot}
    \hat H_{\rm tot}=\hat H_0+\hat H_N+\hat H_S+\hat H_T,\\
\end{equation}
which are defined in detail below.
The core Hamiltonian $\hat{H}_0$ accounts for the QCR charging energy and the qubit. 
In the appropriate basis,\cite{hsu_2020} it assumes the form
\begin{equation}
    \hat H_0  
    %= \hat U \hat H_0 \hat U^\dagger
    =\frac{1}{2C_N}\hat Q_N^2+\hat H_{\varphi},
\end{equation}
where $C_N =C_c+C_{\Sigma m}$ is the total normal-metal capacitance with $C_c$ the coupling capacitance and $C_{\Sigma m}=C_m+C_j$ the junction capacitance, and $\hat Q_N$ denotes the charge on the normal-metal island. The Hamiltonian $\hat{H}_{\varphi}$ includes the charging, Josephson, and inductive energies of the qubit. 
The normal-metal island $\hat H_N$ and the superconducting lead $\hat H_S$ Hamiltonians 
are given by
\begin{align}
\hat H_N = & \sum\limits_{l\sigma} \varepsilon_l \hat d^\dagger_{l\sigma} \hat d^{}_{l\sigma}, \\
\hat H_S = & \sum\limits_{k\sigma} \epsilon_k \hat c^\dagger_{k\sigma} \hat c^{}_{k\sigma}+ \sum\limits_{k} (\tilde \Delta_k \hat c^\dagger_{k\uparrow} \hat c^\dagger_{-k\downarrow} + \textrm{H.c.} ),
\end{align}
where $l$ and $k$ label the eigenstates for electrons with energy $\varepsilon_l$ and $\epsilon_k$ in the normal metal and the superconductor, respectively, $\hat c_{k\sigma}$ and $\hat d_{l\sigma}$ are the corresponding annihilation operators with spin $\sigma$, and $\tilde \Delta_k=\Delta_k e^{-i \frac{2e}{\hbar}Vt}$ is the gap parameter after a time-dependent unitary transformation $\hat U_V (t)=\Pi_{k\sigma }\exp(i\frac{e}{\hbar}Vt \hat c_{k\sigma }^\dag  \hat c^{}_{k\sigma})$, with $V$ the bias voltage and $e$ the elementary charge.
In the chosen basis, the tunneling Hamiltonian $\hat H_T$ captures also the QCR-qubit coupling,
\begin{equation}
    \hat H_T = \sum_{kl\sigma} T_{lk} \hat{d}^\dagger_{l\sigma} \hat{c}_{k\sigma} e^{-i\frac{e}{\hbar}\left(\hat{\phi}_N -  Vt\right)} e^{-i\frac{e}{\hbar} \alpha \hat \phi_0}+ \mathrm{H.c.},
\end{equation}
where $\alpha = C_c/C_N$, $\hat{\phi}_0$ is the phase of the qubit node coupled to the QCR, $T_{lk}$ is the tunneling matrix element, and H.c. denotes the Hermitian conjugate.

For a highly anharmonic qubit, we can project the total Hamiltonian $\hat H_{\rm tot}$ onto the qubit subspace $\ket{0}$ and $\ket{1}$. The core Hamiltonian becomes 
\begin{equation}\label{eq:H0}
    \hat H_0 =\frac{1}{2C_N}\hat Q_N^2+\frac{\hbar\omega_{10}}{2}\hat\sigma^z,
\end{equation}
where $\omega_{10}$ is the qubit frequency and $\hat\sigma^z$ is the Pauli Z matrix. The projected tunneling Hamiltonian reads
\begin{align}
    \hat H_T &=\sum_{kl\sigma} T_{lk} \hat{d}^\dagger_{l\sigma} \hat{c}_{k\sigma}e^{-i\frac{e}{\hbar}\left(\hat{\phi}_N -  Vt\right)}\notag \\
    &\times \left[A^d\hat\sigma^z+A^r(\hat\sigma^+ +\hat\sigma^-)+A^f \hat I \right]+\mathrm{H.c.},
    \label{eq:H_T_QCR}
\end{align}
where $\hat\sigma^{\pm}$ and $\hat I$ are the other Pauli matrices and the coefficients
\begin{equation}
    A^d=\frac{1}{2}(M_{11}-M_{00}), \
    A^r=M_{10}, \ A^f=\frac{1}{2}(M_{11}+M_{00}),
\end{equation}
are given in terms of the matrix elements $M_{mm'}=\bra{m'} e^{-i\frac{e}{\hbar} \alpha \hat\phi_0} \ket{m}$. In general, the matrix element can also depend on the charge of the QCR, not only on the qubit states; we disregard this dependence for simplicity and, for consistency, we neglect charge dispersion in the qubit spectrum as well. For more on the matrix elements, we refer the reader to Ref.~\onlinecite{hsu_2020}, where explicit equations for the matrix elements of transmon and C-shunted flux qubits were derived in the regime of weak anharmonicity, $|\omega_{21}/\omega_{10} -1| \ll 1$, where $\omega_{21}$ is the transition frequency between the second and the first excited states. Note that after this projection, we only consider tunneling events accompanied by an exchange of either no photons or a single photon. This approach is in general not appropriate for harmonic or weakly anharmonic qubits, although it can be a good approximation for low-impedance qubits or weak qubit-QCR coupling.\cite{hsu_2020}

\subsection{Master equation}
\label{sec:QCR_master}

In the weak tunneling regime, the tunneling Hamiltonian $\hat{H}_T$ can be treated as a perturbation, and to calculate transition rates, a Fermi golden rule approach can be used.\cite{Silveri_2017,hsu_2020} To study the dynamics of the QCR, we derive instead a master equation for the reduced density matrix $\hat{\rho}(q)$. If not needed below, we do not explicitly express the dependence on time $t$. More precisely, we assume the electronic degrees of freedom in the normal island and superconducting lead to be in thermal equilibrium at temperature $T_N$, we take the trace of the total density matrix $\hat{\rho}_t$ over them, and project the result onto the eigenstates of the normal-metal charge $\ket{q}$, $q=0,\,\pm1,\,\pm2,\ldots$, with $\hat Q_N\ket{q}=eq\ket{q}$: $\hat{\rho}(q,q') = \bra{q} \mathrm{Tr}_{\hat{c},\hat{d}} \hat{\rho}_t \ket{q'}$, with $\mathrm{Tr}_{\hat{c},\hat{d}}$ the trace over the superconducting ($\hat{c}$) and normal ($\hat{d}$) electron subsystems.
Since we derive the master equation within the Born--Markov and secular approximations,
the equations of motion for the diagonal part form a closed set.\cite{BreuerPetruccione} Here, we focus only on the part diagonal in the charge degree of freedom of the normal-metal island, $\hat{\rho}(q)\equiv \hat{\rho}(q,q)$. Thus,
for each charge value, the reduced density matrix is a $2\times2$ matrix that can be decomposed in terms of Pauli matrices, $
    \rho_\mu(q) = \mathrm{Tr} \left[\hat{\rho}(q) \hat{\sigma}^\mu\right]
$
with $\mu= z, \pm, I$ and $\rho_-(q) = \rho_+(q)^*$.

The equations of motion for the reduced density matrix follows from the von Neumann equation,
\begin{align}
\frac{d}{dt}\rho_z(q) =&-i\mathrm{Tr} \{[\hat H_T,\hat \rho_t]\hat \sigma^z \ket{q} \bra{q}\}\!=-i\langle\langle[\hat \sigma^z\ket{q} \bra{q},\hat H_T]\rangle\rangle, \label{eq:rhozq_EOM} \\
\frac{d}{dt}\rho_+(q) =&\;i\omega_{10}\rho_+(q) -i\mathrm{Tr} \{[\hat H_T,\hat \rho_t]\hat \sigma^+ \ket{q} \bra{q}\} \notag\\
=& i\omega_{10}\rho_+(q) -i\langle\langle[\hat \sigma^+\ket{q} \bra{q},\hat H_T]\rangle\rangle, \label{eq:rho+q_EOM} \\
\frac{d}{dt}\rho_I(q) =&-i\mathrm{Tr} \{[\hat H_T,\hat \rho_t]\hat I \ket{q} \bra{q}\}=-i\langle\langle[\hat I\ket{q} \bra{q},\hat H_T]\rangle\rangle, \label{eq:rhoIq_EOM}
\end{align}
where $\langle\langle\,\cdot\,\rangle\rangle$ denotes averaging with respect to the total density matrix.
We can find the averages in the right-hand sides of Eqs.~\eqref{eq:rhozq_EOM}--\eqref{eq:rhoIq_EOM} by solving the equations governing their temporal evolution within the standard Born--Markov-secular approximation scheme. The procedure is described in Appendix~A of Ref.~\onlinecite{catelani_2012}, and hence here we present only the final form of the master equation for the reduced density matrix. For the diagonal part of the reduced density matrix we have
\begin{align}
    \frac{d}{dt}\rho_0(q)&=\Gamma^+_{q-1,00}\rho_0(q-1)+\Gamma^-_{q+1,00}\rho_0(q+1) \notag \\
    &+\Gamma^+_{q-1,10}\rho_1(q-1)+\Gamma^-_{q+1,10}\rho_1(q+1)\notag \\
    &-(\Gamma^+_{q,01}+\Gamma^-_{q,01}+\Gamma^+_{q,00}+\Gamma^-_{q,00})\rho_0(q) \label{eq:rho0}\\
    \frac{d}{dt}\rho_1(q)&=\Gamma^+_{q-1,11}\rho_1(q-1)+\Gamma^-_{q+1,11}\rho_1(q+1)\notag \\
    &+\Gamma^+_{q-1,01}\rho_0(q-1)+\Gamma^-_{q+1,01}\rho_0(q+1) \notag \\
    &-(\Gamma^+_{q,10}+\Gamma^-_{q,10}+\Gamma^+_{q,11}+\Gamma^-_{q,11})\rho_1(q), \label{eq:rho1}
\end{align}
where $\rho_0(q)=[\rho_I(q)-\rho_z(q)]/2$, $\rho_1(q)=[\rho_I(q)+\rho_z(q)]/2$, and the transition rates are
\begin{equation}\label{eq:rate_qmmp}
    \Gamma^{\pm}_{q,mm'}(V)=\frac{R_K}{R_T} M^2_{mm'}\sum\limits_{\tau=\pm 1} F(\tau eV+\hbar\omega_{mm'}-E^{\pm}_{q}),
\end{equation}
where the superscript $\pm$ denotes whether the initial charge $q$ increases or decreases by one electron charge. Here $R_K=h/e^2=25.813~\textrm{k}\Omega$ is the von Klitzing constant, $R_T$ is the tunneling resistance of the SIN junction, $M^2_{mm'}\equiv |M_{mm'}|^2$ is used as a short-hand notation, $\omega_{mm}=0$, $\omega_{01}=-\omega_{10}$, and $E_q^{\pm}=E_N (1\pm 2q)$ with $E_N=e^2/(2C_N)$ the charging energy of the normal-metal island. The normalized rate of single-electron tunneling is defined as
\begin{equation}\label{eq:FE_def}
    F(E)=\frac{1}{h}\int d\epsilon \; n_S(\epsilon) f(\epsilon-E) [1-f(\epsilon)],
\end{equation}
where $f$ denotes the Fermi--Dirac distribution function at temperature $T_N$. The normalized superconductor density of states is of the form
\begin{equation}
n_{S}(\varepsilon) = \left|\mathrm{Re} \left[ \frac{\varepsilon + i \gamma_{D}\Delta}{\sqrt{(\varepsilon + i \gamma_{D}\Delta)^{2} - \Delta^{2}}}\right] \right|,
\label{eq:ns}
\end{equation}
where $\gamma_D$ is the Dynes parameter which accounts for a possible broadening of density of states. The quantities introduced in Eqs.~(\ref{eq:rate_qmmp})--(\ref{eq:ns}) are analogous to those of Refs.~\onlinecite{Silveri_2017,hsu_2020}.

The equation for the off-diagonal part of the reduced density matrix $\rho_+(q)$ is
\begin{align}
  & \frac{d}{dt}\rho_+(q) =i[\omega_{10}+\delta\omega(q)] \rho_+(q) \label{eq:rho+} \\
  & +  \frac{R_K}{R_T}\left(|A^f|^2-|A^d|^2\right)\sum_{\tau,\eta=\pm}F(\tau eV+E^\eta_q)\rho_+(q+\eta 1) \nonumber \\
  & -  \frac{R_K}{R_T}\left(|A^f|^2+|A^d|^2\right)\sum_{\tau,\eta=\pm}F(\tau eV-E^\eta_q)\rho_+(q) \nonumber \\
  &  - \frac12  \frac{R_K}{R_T}|A^r|^2 \sum_{\tau,\eta,\nu=\pm}F(\tau eV + \nu \omega_{10}-E^\eta_q)\rho_+(q)
  \nonumber
\end{align}
where the charge-dependent frequency shift
\begin{align}
    & \delta\omega(q) = \mathrm{PV} \int\frac{d\omega}{2\pi} \label{eq:shifts} \\ 
    & \bigg\{\frac{1}{\omega} \left[ \Gamma^+_{q,00}(\omega)+\Gamma^-_{q,00}(\omega)-\Gamma^+_{q,11}(\omega)-\Gamma^-_{q,11}(\omega) \right] \nonumber \\
    & +\left(\frac{1}{\omega+\omega_{10}}-\frac{1}{\omega-\omega_{10}}\right) \left[ \Gamma^+_{q,10}(\omega)+\Gamma^-_{q,10}(\omega)\right] \bigg\} \nonumber
\end{align}
accounts for the second-order perturbation to the qubit energy levels due to the interaction with the QCR; here PV denotes the Cauchy principal value and the rates $\Gamma^\pm_{q,mm'}(\omega)$ are defined as in Eq.~\eqref{eq:rate_qmmp} but with the replacement $\omega_{mm'} \to \omega$ on the right side. As noted in Ref.~\onlinecite{catelani_2012}, such an expression for the frequency shift is not appropriate for weakly anharmonic systems, like those considered in Ref.~\onlinecite{hsu_2020}, since it does not account for the level repulsion originating from higher levels. The QCR-induced frequency shift for a harmonic system, a superconducting resonator, was measured recently.\cite{Silveri_2019} Since for typical experimental parameters we have $\delta\omega(q) \ll \omega_{10}$ (cf. Ref.~\onlinecite{hsu_2020}), we henceforth do not consider the frequency shift further.

The assumption that the charge degree of freedom can be treated as a bath under the standard Born--Markov approximation can be expressed mathematically by assuming that the components of the density matrix can be factorized, $\rho_m(q,t) = \rho_m(t) \bar{\rho}(q)$. Then one can trace out the charge (\textit{i.e.}, sum over $q$) in Eqs.~(\ref{eq:rho0}), (\ref{eq:rho1}), and (\ref{eq:rho+}) to find
\begin{align}\label{eq:master_qaverage}
    \frac{d}{dt}\rho_0 & = \Gamma_{10} \rho_1 - \Gamma_{01}\rho_0 \, , \\
    \frac{d}{dt}\rho_1 & = \Gamma_{01} \rho_0 - \Gamma_{10}\rho_1\, , \\
    \frac{d}{dt}\rho_+ & = i\omega_{10} \rho_+ -
    \frac{1}{2}\left(\Gamma_{10} +\Gamma_{01}\right)\rho_+ \label{eq:rho+q}\\ & -\frac{1}{2}\Gamma_{00}\left(1-\frac{M_{11}}{M_{00}}\right)^2 \rho_+ \, , \nonumber
\end{align}
with
\begin{equation}\label{eq:rate_charge_avg}
    \Gamma_{mm'} = \sum_q \left(\Gamma^+_{q,mm'} + \Gamma^-_{q,mm'}\right) \bar{\rho}(q)
\end{equation}
denoting the charge-averaged transition rates.
The last two terms in Eq.~(\ref{eq:rho+q}) give the qubit decoherence rate $1/T_2$ in the usual combination $1/(2T_1) + 1/T_\varphi$.
The pure dephasing term $\propto\Gamma_{00}$ is actually a lower limit on the QCR-induced pure dephasing, appropriate if the qubit charge dispersion can be neglected\cite{catelani_2014} as assumed here. See also Ref.~\onlinecite{hsu_2020} for estimates that take the charge dispersion into account.

In Ref.~\onlinecite{Silveri_2017} it was argued that one can expect factorization when the quasi-elastic rates $\Gamma_{ii}$ are fast compared to the inelastic transitions $\Gamma_{ij}$, $j\neq i$, and that one can further simplify the expression for the rates $\Gamma_{ij}$  in Eq.~(\ref{eq:rate_charge_avg}) by noting that subleading terms in the Taylor expansion around $q=0$ can be neglected when the charging energy $E_N$ is the smallest energy scale in the problem (except for $\gamma_D \Delta$), $\Gamma_{mm'} \approx \Gamma^+_{0,mm'} + \Gamma^-_{0,mm'}$. By solving the two above master equations, we can reconsider if the factorization assumption holds, and also when the approximation for the rates is viable.

\subsection{Approximate solution to the master equation}\label{sec:sol_master_QCR}

Finding an analytical solution to the master equation Eqs.~\eqref{eq:rho0}, \eqref{eq:rho1} and \eqref{eq:rho+} is clearly only possible in an approximate way. In fact, the approximation just discussed for the rates $\Gamma_{mm'}$ hints at a possible approach, and our first step is to expand the rates $\Gamma^\pm_{q,mm'}$ as functions of the charging energy $E_N$. Concretely, for the quasi-elastic rates $\Gamma^{\pm}_{q,mm}$ we find at first order
\begin{equation}\label{eq:el_rate_exp}
    \Gamma^{\pm}_{q,mm} \approx \frac{R_K}{R_T}M_{mm}^2\sum_{\tau=\pm1}F(\tau eV) \left[1+b(eV,0) (1\pm 2q)\right]\, ,
\end{equation}
where
\begin{equation}
    b(E,\omega) = E_N \frac{F'(E+\hbar\omega)+F'(-E+\hbar\omega)}{F(E+\hbar\omega)+F(-E+\hbar\omega)} \,.
   \label{eq:b}
\end{equation}
A necessary condition for neglecting higher-order corrections (see also Appendix~\ref{app:2nd}) is $b\ll 1$. Since from the identity
\begin{align}
    F'(E)=\frac{1}{h}\int d\epsilon \; n_S(\epsilon)  f(\epsilon-E) 
    [1-f(\epsilon)] \\ \times \frac{1}{k_B T_N}\frac{e^{(\epsilon-E)/(k_B T_N)}}{1+e^{(\epsilon-E)/(k_B T_N)}}  \nonumber
\end{align}
it follows that $F'(E)<F(E)/(k_B T_N)$, we conclude that $b<E_N/(k_B T_N)$ and the necessary conditions becomes $E_N/(k_B T_N) \ll 1$. A more careful analysis based on the equations presented in Appendix~A of Ref.~\onlinecite{hsu_2020} shows that the inequality is saturated [$b(E,0)\approx E_N/(k_B T_N)$] in the so-called thermal activation regime, $E_\mathrm{co} < E \lesssim \Delta$, with 
\begin{align}
E_\mathrm{co} \approx \Delta- k_B T_N \bigg\{ &\ln \left( \frac{\sqrt{\pi}k_B T_N}{\gamma_D \Delta}\right)\notag \\&  + \frac{1}{2} \ln \left[ \ln \left( \frac{\sqrt{\pi}k_B T_N}{\gamma_D \Delta} \right)\right]\bigg \} .
\label{eq:Eco}
\end{align}
At lower energies, $k_BT_N \lesssim E < E_\mathrm{co}$, we have $b(E,0)\approx E_N/E$, and  for $E\ll k_B T_N$ we have $b(E,0) \approx E_N/(2k_B T_N)$. At higher energies, $E\gg\Delta$, we find again $b(E,0)\approx E_N/E$.

Expansions similar to that in Eq.~(\ref{eq:el_rate_exp}) can also be performed for the decay $\Gamma^\pm_{q,10}$ and excitation $\Gamma^\pm_{q,01}$ rates; they can be obtained by the replacements $F(\tau eV)\to F(\tau eV \pm \hbar\omega_{10})$, $b(eV,0) \to b(eV,\pm\omega_{10})$ in Eq.~(\ref{eq:el_rate_exp}), with positive (negative) sign for the decay (excitation) rate. Focusing on the case $V=0$, we have $b(0,\omega_{10}) \approx E_N/(\hbar\omega_{10})$, and since qubits are usually operated at temperatures $k_B T_N\ll \hbar\omega_{10}$ we find that the decay rate has a relatively much weaker dependence on the QCR charge than the elastic rate. Conversely, using the identity $F(-E) = e^{-E/(k_B T_N)}F(E)$, we estimate $b(0,-\omega_{10}) \approx E_N[1/(k_B T_N)-1/(\hbar\omega_{10})]$, and hence the excitation rates has a somewhat stronger dependence on charge in comparison to the elastic rate. As discussed in Ref.~\onlinecite{hsu_2020}, the useful QCR parameters are such that at $V=0$ we expect $\Gamma^\pm_{q,01}\ll \Gamma^\pm_{q,10} \lesssim \Gamma^\pm_{q,mm}$. Therefore we can neglect the charge dependence of the decay and excitation rates: for the decay rate, because its charge dependence is weaker by the small factor $k_B T_N/(\hbar\omega_{10})$; for the excitation rate because the rate itself is already small compared to the other rates. Although we have considered the validity of this approximations only at $V=0$, we can use them also at finite bias thanks to the inequality $b<E_N/(k_B T_N)$; we caution, however, that there are combinations of temperature $T_N$ and bias $V$ for which $\Gamma_{10} \gg \Gamma_{00}$,\cite{hsu_2020} in which cases the results derived below may not be quantitatively correct.

Within the introduced approximations, the master equation assumes the form
\begin{align}
\dot{\rho}_0(q)& =(1-\eta)\big\{\left[1+b(1+2q)\right]\rho_0(q+1) \notag \\
&+\left[1+b(1-2q)\right]\rho_0(q-1)-2\left[1-b\right]\rho_0(q) \big\}\notag \\
&+ \frac12 \Gamma_d \left[\rho_1(q+1)+\rho_1(q-1)\right]-\Gamma_u \rho_0 (q), \label{eq:rho0_1st_expandF}
\\
\dot{\rho}_1(q)&=(1+\eta)\big\{\left[1+b(1+2q)\right]\rho_1(q+1)\notag \\ 
&+ \left[1+b(1-2q)\right]\rho_1(q-1)-2\left[1-b\right]\rho_1(q)\big\} \notag \\ 
&+\frac12 \Gamma_u \left[\rho_0(q+1)+\rho_0(q-1)\right]-\Gamma_d \rho_1 (q), 
\label{eq:rho1_1st_expandF}
\\
\dot{\rho}_+(q) & = i\tilde\omega_{10} \rho_+(q)+ \sqrt{1-\eta^2}\big\{ \left[1+b(1+2q)\right]\rho_+(q+1) \notag \\
&+\left[1+b(1-2q)\right]\rho_+(q-1)\big\} \notag \\ & -2\left[1-b\right]\rho_+(q) -\frac12\left(\Gamma_u+\Gamma_d\right)\rho_+(q), 
\label{eq:rho+_elastic_1st}
\end{align}
where we use the short-hand notation $b=b(eV,0)$ and we have rescaled the time variable by half of the average elastic transition rate,
\begin{equation}
    t\to \Gamma_{el} t \, , \quad \Gamma_{el}(eV) = \frac{R_K}{R_T}\frac{M^2_{00}+M^2_{11}}{2}\sum_{\tau=\pm1} F(\tau eV).
    \label{eq:Gel}
\end{equation}
The parameter $\eta$ accounts for the difference in the elastic matrix elements, $\eta= (M_{11}^2-M_{00}^2)/(M_{11}^2+M_{00}^2)$, and is usually small $\eta \ll 1$. We will neglect $\eta$ hereinafter; it can be treated perturbatively (see Appendix~\ref{app:charge_correction}). The up/down rates are the normalized excitation and decay rates: $\Gamma_u = (\Gamma^{+}_{q,01}+\Gamma^{-}_{q,01})/\Gamma_{el}$ and $\Gamma_d = (\Gamma^{+}_{q,10}+\Gamma^{-}_{q,10})/\Gamma_{el}$, where in our approximation the rates $\Gamma^{\pm}_{q,mm'}$ are calculated at zeroth order in $E_N/(k_B T_N)$ and are therefore independent of charge $q$. Similarly, $\tilde\omega_{10} = \omega_{10}/\Gamma_{el}$ is the normalized qubit frequency.

Focusing first on the coupled equations (\ref{eq:rho0_1st_expandF}) and (\ref{eq:rho1_1st_expandF}), we look for solutions decaying exponentially in time, $\rho_{mj}(q,t) = e^{-\lambda_j t}\rho_{mj}(q)$. We find that further progress can be made by distinguishing ``charge-type'' solutions from ``qubit-type'' solution; linear combinations of solutions of the first type are such that the qubit polarization is time-independent, which explains the chosen nomenclature. The charge-type solutions can be found by making the Ansatz
\begin{align}\label{eq:charge_ans}
    \rho_{0j}(q) & = \frac{\Gamma_d}{\Gamma_u+\Gamma_d}\left[\bar\rho_j(q) + \delta\bar\rho_j(q)\right], \\
    \rho_{1j}(q) & = \frac{\Gamma_u}{\Gamma_u+\Gamma_d}\left[\bar\rho_j(q) - \delta\bar\rho_j(q)\right]. \nonumber
\end{align}
As discussed in Appendix~\ref{app:charge_correction}, assuming $\gamma = \Gamma_d-\Gamma_u \ll 1$ implies that $\delta\bar\rho_j$ is a small correction that can be neglected; here we note that this condition can be satisfied not only when the up/down rates are similar, but also when they are both small compared to the elastic rates.  Under this assumption the equation for $\bar\rho_j$ is
\begin{align}\label{eq:barrhoj_bf_expand}
    -\lambda_j \bar\rho_j =& \left(1+b+\bar{\Gamma}/2\right)\left[\bar\rho_j(q+1)+\bar\rho_j(q-1)-2\bar\rho_j(q)\right] \nonumber \\
    & +2bq\left[\bar\rho_j(q+1) -\bar\rho_j(q-1)\right] + 4b\bar\rho_j(q)
\end{align}
with $\bar\Gamma= (\Gamma_u+\Gamma_d)/2$.
To find an approximate solution to this finite difference equation, we convert it into a differential equation by treating $q$ as a continuous variable and Taylor expanding $\bar\rho(q\pm 1)$ up to second order,
\begin{equation}\label{eq:Taylor}
    \bar\rho_{j}(q\pm1)\approx\bar\rho_{j}(q)\pm\partial_q\bar\rho_{j}(q)+\frac{1}{2}\partial^2_q\bar\rho_{j}(q).
\end{equation}
After the identification of the derivative with the momentum operator $p=-i\partial_q$ conjugate to $q$, and the canonical transformation $p=\tilde{p}+2ibq/(1+b+\bar\Gamma/2)$, the equation assumes the form of the time-independent Schr\"odinger equation for an harmonic oscillator,
\begin{equation}\label{eq:barrhoj}
    \lambda_j \bar\rho_j = \left[\left(1+b+\bar\Gamma/2\right)\tilde{p}^2 +\frac{4b^2}{1+b+\bar\Gamma/2}q^2 -2b \right] \bar\rho_j \,.
\end{equation}
Therefore, we can immediately find the decay rates as $\lambda_j = 4bj$, $j=0,\,1,\,2,\ldots$ For the stationary state $j=0$, the distribution function is Gaussian,
\begin{equation}
    \bar\rho_0 (q) = \frac{1}{Z} e^{-2bq^2}
\end{equation}
where $Z$ is a normalization constant (see also Appendix~\ref{app:charge_correction}); by comparing this expression to the standard form $e^{-\mathcal{H}/(k_B T_Q)}/Z$ with $\mathcal{H}=E_Nq^2$ the normal-island charging energy and $T_Q$ an effective temperature for the charge, we find $k_B T_Q = E_N/(2b)$; in equilibrium, $b(eV=0) = E_N/(2k_B T_N)$ and hence $T_Q = T_N$. We have also verified that this definition reproduces the numerical results for finite bias presented in the Appendix of Ref.~\onlinecite{Silveri_2017}.

For the qubit-type modes we start from the Ansatz
\begin{equation}\label{eq:q_ans}
    \rho_{0j}(q) = \tilde\rho_j(q) + \delta\tilde\rho_j(q), \quad
    \rho_{1j}(q) = -\tilde\rho_j(q) + \delta\tilde\rho_j(q). \
\end{equation}
Assuming again $\gamma \ll 1$, we neglect $\delta\tilde\rho_j$, and the equation for $\tilde\rho_j$ is thus given by
\begin{align}
    -&\lambda_j \tilde\rho_j = \left(1+b-\bar{\Gamma}/2\right)\left[\tilde\rho_j(q+1)+\tilde\rho_j(q-1)-2\tilde\rho_j(q)\right] \nonumber \\
    & +2bq\left[\bar\rho_j(q+1) -\bar\rho_j(q-1)\right] + \left(4b-2\bar\Gamma\right)\bar\rho_j(q)
\label{eq:rhot}
\end{align}
Proceeding as described above (Taylor expansion plus canonical transformation, leading to harmonic oscillator equation), we arrive at $\lambda_j = 4bj + \Gamma_u + \Gamma_d$. Therefore, for each charge-type mode there exist a corresponding qubit-type mode that decays faster by the rate $2\bar\Gamma=\Gamma_u+\Gamma_d$. However, the charge dependencies of the modes are identical, $\tilde\rho_j(q) = \bar\rho_j(q)$, under the condition $\bar\Gamma \ll 1$ (see Appendix~\ref{app:charge_correction}).

Equation (\ref{eq:rho+_elastic_1st}) can also be solved in a similar manner, and looking for solutions of the form $\rho_+(t) = e^{(i\tilde\omega_{10} - \lambda_j)t}\rho_{+j}(q)$, we find $\lambda_j = 4bj + \bar\Gamma$, with the last terms being as usual half the (excess) decay of the qubit-type modes. Here the assumption $\gamma\ll 1$ is not needed; note that by neglecting $\eta$, we are neglecting a small pure dephasing contribution analogous to that discussed below Eq.~(\ref{eq:rate_charge_avg}).

Within the approximations used, the above results imply that at leading order the full reduced density matrix $\hat\rho$ can be written as
\begin{equation}\label{eq:rhofactor}
    \hat{\rho}(q,t) = \rho_2(t) \sum_{j=0} \alpha_j \bar\rho_j(q) e^{-4bj \Gamma_{el} t}
\end{equation}
where $\rho_2(t)$ is a 2$\times$2 matrix describing the usual temporal evolution of the qubit over the time scale defined by $T_1=1/(2\bar{\Gamma})$, and the coefficients $\alpha_j$ are determined by the initial conditions (together with normalization); here we have restored dimensionful time units by undoing the rescaling in Eq.~(\ref{eq:Gel}). The relaxation of the charge takes place over the time scale $\tau_Q = 1/\Gamma_Q$ where
\begin{equation}\label{eq:GammaQ}
    \Gamma_Q (V) = 4b(eV,0) \Gamma_{el}(eV)
\end{equation}
The time $\tau_Q$ represents the non-equilibrium generalization of the QCR's $RC$-time; indeed, at equilibrium ($V=0$) and assuming $M_{00}^2+M_{11}^2 \simeq 2$ (cf. Table \ref{tab:parameters_QCR2level}) we find $\tau_Q =R_T C_N/(2\gamma_D)$, with the Dynes parameter and the factor 2 accounting for the high sub-gap resistance of the two contacts.
For $\tau_Q \ll T_1$, we can neglect all terms with $j>0$, and we arrive at the factorization discussed above Eq.~(\ref{eq:master_qaverage}). We note that since $b\ll 1$, this requirement is more stringent than simply asking for the elastic rates to be faster than the inelastic ones ($1/\Gamma_{el} < T_1$). This inequality is satisfied for the typical parameters considered in Ref.~\onlinecite{hsu_2020}, but in general the condition $\tau_Q \ll T_1$ is violated. However, we now argue that this does not affect the theoretical estimates for the rates of the previous works.\cite{hsu_2020,Silveri_2017}

\begin{figure}[tb]
  \centering
  \includegraphics[width=1.0\linewidth]{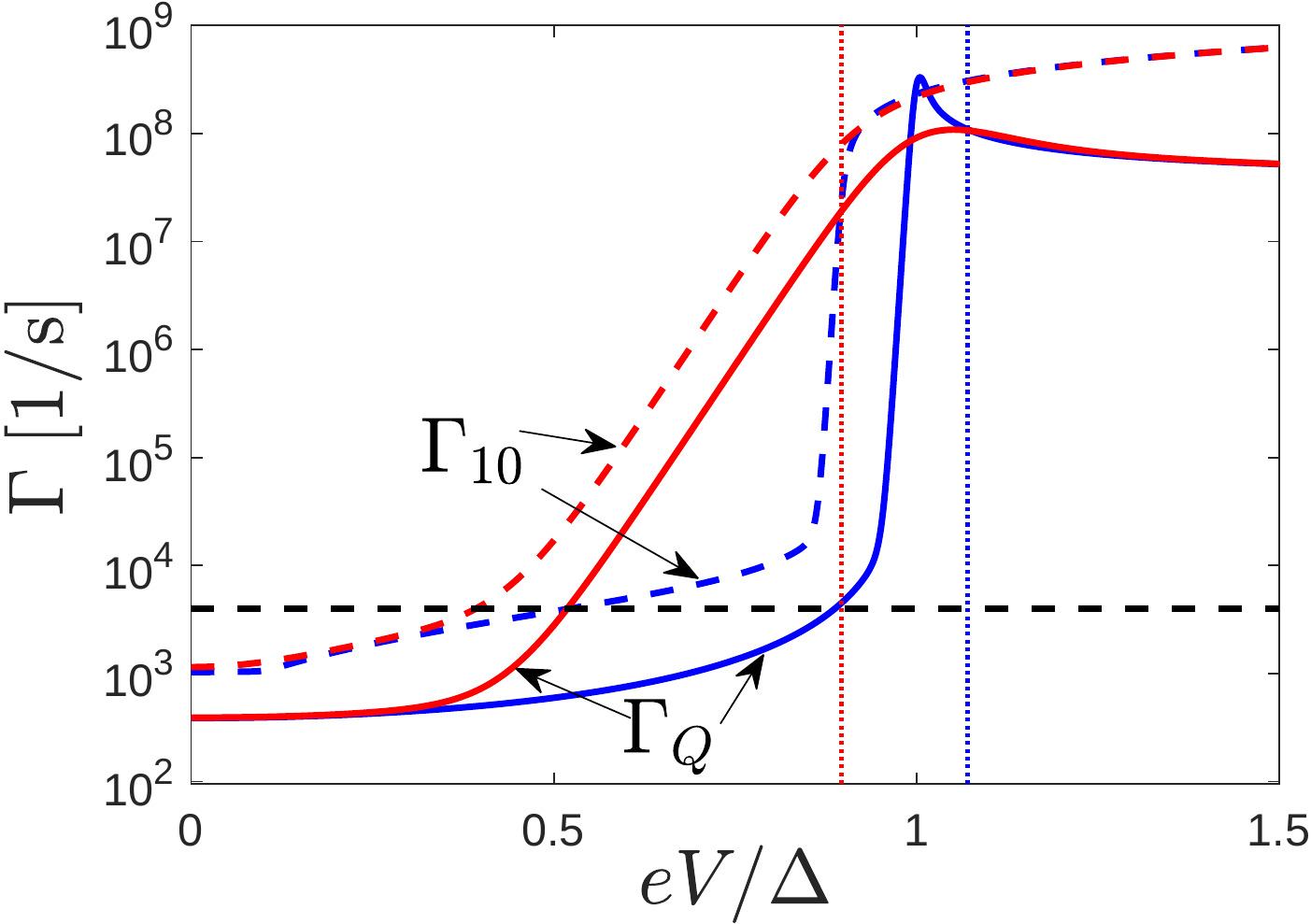}
  \caption{\label{fig:Gamma_charge_re}Charge relaxation rate $\Gamma_Q$ of Eq.~(\ref{eq:GammaQ}) (solid lines) and QCR-induced qubit decay rate $\Gamma_{10}$ (dashed lines) as functions of normalized bias voltage $eV/\Delta$ for different normal-metal electron temperatures $k_B T_N/\Delta=0.043$ (red) and $k_B T_N/\Delta=0.0043$ (blue). The black dashed horizontal line gives the qubit bare relaxation rate $1/(250~\mu s)$.\cite{serniak_2019} 
  The red (blue) vertical dotted line marks the on-voltage $eV_{\rm on}=\Delta-\hbar\omega_{10}$ ($eV_{\rm on}=E_{\rm co}+\hbar\omega_{10}$) for $k_B T_N/\Delta=0.043$ ($k_B T_N/\Delta=0.0043$). Parameters are given in Table~\ref{tab:parameters_QCR2level}.}
\end{figure}

Let us assume that to begin with, we wait for a sufficiently long time $t\gg \tau_Q$, so that we start from the equilibrium state at $V=0$. Manipulation of the qubit, e.g., by microwave drives, affects the qubit part $\rho_2$ of the density matrix, but not the charge part, which therefore remains $\bar\rho_0(q)$. When the qubit needs to be reset, the bias voltage is quickly increased to the operating point $V_\mathrm{on}$ at which the qubit dynamics is dominated by the QCR-induced decay rate (at $V=0$ non-QCR mechanism are dominating, so that the QCR does not affect the qubit). The value of $V_\mathrm{on}$ depends on device parameters and also on the temperature regime (high or low, as defined in Ref.~\onlinecite{hsu_2020}); for our purposes, one only needs to know that in the high-temperature thermal activation case $b(eV_\mathrm{on},0)\approx E_N/(k_B T_N)$ and in the low-temperature one $b(eV_\mathrm{on},0)\approx E_N/\Delta$. With these expression, one can check that at $V_\mathrm{on}$ the condition $\Gamma_{Q}(V_\mathrm{on}) \ll \Gamma_{10}^\mathrm{on}$ holds, see Fig.~\ref{fig:Gamma_charge_re}. Since the QCR is kept at $V_\mathrm{on}$ for a time of order $1/\Gamma_{10}^\mathrm{on}$, the charge distribution does not change much over that time, and the factorized form of the density matrix, $\hat\rho(q,t) = \hat\rho_2(t)\bar\rho_0(q)$ can be used at all times, where the charge-dependent factor is always the one in equilibrium. Calculating the charge-averaged transition rates  as defined in Eq.~(\ref{eq:rate_charge_avg}), we remind that because of the symmetry of $\bar\rho_0(q)$, the corrections due to the finite charging energy are in fact decreased by the factor $b^2 \approx [E_N/(k_B T_N)]^2$.

\renewcommand{\arraystretch}{1.25}
\begin{table}[bt]
    \centering
    \begin{tabular}{c c c c c c c}
         \multicolumn{1}{c|}{} &  $C_c$ & $C_N$ & $R_T$ & $\gamma_D$ & $\Delta$ \\
         \cline{1-6}
         \multicolumn{1}{c|}{QCR} & \SI{5}{\femto\farad} & \SI{15}{\femto\farad} & \SI{50}{\kilo\ohm} & \num{e-5} & \SI{200}{\micro\electronvolt} \\
         \\[-0.2cm]
        \multicolumn{1}{c|}{} & $\omega_{10}/(2\pi)$ & $M^2_{00}$ & $M^2_{11}$ & $M^2_{10}$ \\
         \cline{1-5}
        \multicolumn{1}{c|}{qubit} & \SI{5}{\giga\hertz} & 0.99 & 0.97 & 0.01 \\
         \\[-0.2cm]
    \end{tabular}
    \caption{Parameter values used in  numerical calculations. The matrix elements $M^2_{mm'}$ are evaluated numerically (see Ref.~\onlinecite{hsu_2020}) for an offset-charge-sensitive transmon\cite{serniak_2019} with $E_J/E_C=20$; with this choice, the qubit has higher anharmonicity than for typically larger $E_J/E_C$ ratio and also higher matrix element $M_{01}$ at given $\alpha$. To counteract this latter increase, we choose a lower coupling capacitance $C_c$ (and hence total normal-metal capacitance $C_N$) than in Ref.~\onlinecite{hsu_2020}, so that the QCR-induced qubit decay rate at zero bias is smaller than the bare qubit relaxation rate.}
    \label{tab:parameters_QCR2level}
\end{table}

In the above considerations, we have explicitly considered only the qubit decoherence caused by the QCR itself. However, the working principle of the QCR assumes that at zero bias other mechanisms, not the QCR, are the major source of decoherence. Such mechanisms can be included in the master equation and, as discussed in Appendix~\ref{app:totdec}, only affect the factor $\rho_2(t)$ in Eq.~(\ref{eq:rhofactor}), so that the qubit decoherence rates are given by the sum of the rates due to the QCR and the other mechanisms. Next, we use the factorized form to analyze an alternative qubit reset protocol.

\subsection{Ac control of the QCR}\label{sec:acdrive}

Above and in the previous works,\cite{hsu_2020,Silveri_2017} we have only considered the QCR to be controlled by a dc voltage bias $V$. We employed $V=0$ for the off-state, in which the QCR does not affect the qubit, and a voltage of roughly $\Delta/e$ for the on-state, see Fig.~\ref{fig:Gamma_charge_re}.
In this section, we study another way to operate the QCR. Instead of tuning the dc bias back and forth, we fix it at $eV_{\rm off} = E_{\rm co}-\hbar \omega_{\rm 10}$, the bias at which the QCR-induced qubit relaxation rate begins to increase exponentially with the bias voltage. To turn on the QCR, we add an ac voltage drive of frequency $\Omega_\textrm{AC}$ and amplitude amplitude $V_{\rm AC}$ such that the
total time-dependent voltage is
\begin{equation}
    V(t)=V_{\rm off}+V_{\rm AC}\cos\Omega_{\rm AC}t.
    \label{eq:V_include_AC}
\end{equation}
This kind of rf controlled QCR was coined in Ref.~\onlinecite{viitanen} as rf QCR and it was shown theoretically and experimentally that one can change the QCR-induced dissipation rate by orders of magnitude by changing the rf drive power. However, Ref.~\onlinecite{viitanen} treated the effect quantum mechanically by including the drive mode into the core Hamiltonian of the circuit and using the theory of photon-assisted tunneling of the two-mode system to extract the effective damping rate of the primary mode of interest. In contrast, we consider a classical drive voltage leading to a much more convenient treatment of the problem than that developed in Ref.~\onlinecite{viitanen}.

Consequently in our case, the voltage-dependent phase factor in the tunneling Hamiltonian $\hat{H}_T$, Eq.~(\ref{eq:H_T_QCR}), becomes 
\begin{align}
    &\exp\left[i\frac{e}{\hbar}\int^t_0(V_{\rm off}+V_{\rm AC}\cos\Omega_{\rm AC}t')dt'\right] \notag \\
    &=\exp\left[i\frac{e}{\hbar} \left(V_{\rm off} t+\frac{V_{\rm AC}}{\Omega_{\rm AC}}\sin\Omega_{\rm AC}t\right) \right].
    \label{eq:H_T_phase_AC}
\end{align}
Following the approach of Ref.~\onlinecite{Barone}, we use the Jacobi-Anger expansion to rewrite the phase factor of Eq.~\eqref{eq:H_T_phase_AC} as
\begin{align}\label{eq:H_T_phase_AC_expand}
    & \exp\left[i\frac{e}{\hbar}\left(V_{\rm off}t+\frac{V_\mathrm{AC}}{\Omega_\mathrm{AC}}\sin\Omega_{\rm AC}t\right) \right] = \\ & J_0\left(\frac{eV_{\rm AC}}{\hbar\Omega_{\rm AC}}\right)e^{ieV_{\rm off}t/\hbar} 
    +\sum\limits_{k=1}^\infty J_k\left(\frac{eV_{\rm AC}}{\hbar\Omega_{\rm AC}}\right) \notag \\ & \times\left[ e^{i(k\Omega_\textrm{AC}+eV_{\rm off}/\hbar)t} +(-1)^k e^{-i(k\Omega_\textrm{AC}-eV_{\rm off}/\hbar)t} \right], \notag
\end{align}
where $J_k$ are Bessel functions of the first kind of order $k$.
Using this expansion, we can derive the master equation following similar steps as in Sec.~\ref{sec:QCR_master}. Consequently, the charge-dependent normalized rates $F(\tau eV+\hbar\omega_{mm'}-E^{\pm}_{q})$ are multiplied by the factor $\{J_0[eV_{\rm AC}/(\hbar\Omega_{\rm AC})]\}^2$, and we have additional terms of the form $\{J_k[eV_{\rm AC}/(\hbar\Omega_{\rm AC})]\}^2 F(\tau eV+k\hbar\Omega_{\rm AC}+\hbar\omega_{mm'}-E^{\pm}_{q})$, in which the ac drive introduces an energy shift $k\hbar\Omega_{\rm AC}$. 
Owing to such shifts, the validity of the secular approximation calls for more stringent conditions to be met than in the absence of shifts, but here we simply assume these conditions to be fulfilled, \textit{i.e.}, that there are no significant effective degeneracies introduced by the drive. 

\begin{figure}[tb]
  \centering
  \includegraphics[width=1.0\linewidth]{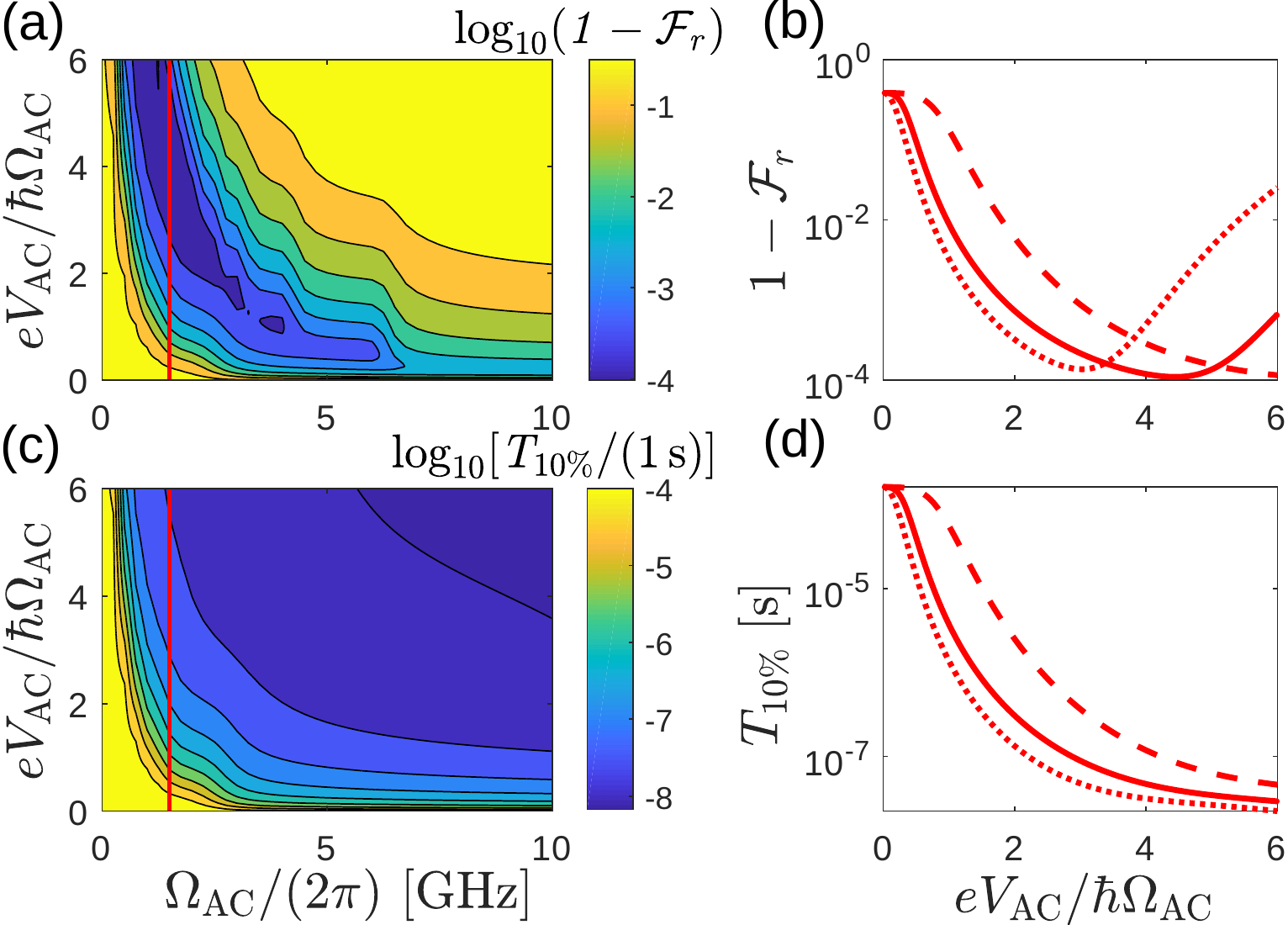}
  \caption{\label{fig:AC_drive}
  (a) Infidelity $1-\mathcal{F}_r$ of Eq.~\eqref{eq:fidelity} and (c) reset time $T_{10\%}$ of Eq.~\eqref{eq:reset_time} as functions of the ratio the ac drive frequency $\Omega_{\rm AC}/(2\pi)$ and and amplitude $V_\textrm{AC}$ for the low-temperature case $T_N/\Delta=0.0043$. We set the off voltage to $eV_{\rm off}=E_{co}-\hbar\omega_{10}$ with $E_{co}$ of Eq.~\eqref{eq:Eco}. We sum the harmonics from $k=-10$ to $k=10$ to arrive at numerically accurate results. Solid lines in panels (b) and (d) show line traces from panels (a) and (c), respectively, as indicated by the red vertical 
 lines at $\Omega_{\rm AC}/(2\pi)=1.5$~GHz. We show also traces for $\Omega_{\rm AC}/(2\pi)=1$~GHz (dashed lines) and $\Omega_{\rm AC}/(2\pi)=2$~GHz (dotted lines) for comparison. The parameter values are given in Table~\ref{tab:parameters_QCR2level}.}
\end{figure}

As discussed in Sec.~\ref{sec:sol_master_QCR}, to calculate the transition rates we can average over the charge distribution; the resulting qubit relaxation rate, including the ac drive, is given by
\begin{align}
   \Gamma_{10} & \approx  \frac{2R_{K}}{R_T}M^2_{10}\sum_{\tau=\pm1}\sum\limits_{k=-\infty}^\infty \bigg\{
    \{J_k[eV_{\rm AC}/(\hbar\Omega_{\rm AC})]\}^2 \notag \\
    & \times F(\tau eV_{\rm off}+k\hbar\Omega_{\rm AC}+\hbar\omega_{10})\bigg\}.
\label{eq:Gamma_AC}
\end{align}
For the qubit excitation rate $\Gamma_{01}$, $+\hbar\omega_{10}$ is replaced with $-\hbar\omega_{10}$ on the right side of Eq.~\eqref{eq:Gamma_AC}.
This result resembles that in Ref.~\onlinecite{viitanen}, where the photon-assisted tunneling processes involving $k$ photons in the drive mode of angular frequency $\Omega_\textrm{AC}$ effectively change the bias voltage by $k\hbar\Omega_\textrm{AC}$ as can be also interpreted from Eq.~\eqref{eq:Gamma_AC}.
As in Ref.~\onlinecite{hsu_2020}, we aim for the fastest possible qubit decay with the highest possible fidelity. Therefore,
we consider here the reset infidelity $1-\mathcal{F}_r$ where, assuming that the QCR-induced excitation is the main contributor to the excitation rate, the fidelity is defined by
\begin{equation}
    \mathcal{F}_r=\frac{\Gamma_{01}}{\Gamma_{10}} \, ,
    \label{eq:fidelity}
\end{equation}
and the reset time
\begin{equation}
    T_{10\%}=\frac{\ln(10)}{\Gamma_{10}} \,. 
    \label{eq:reset_time}
\end{equation}
In the absence of the ac drive, $V_{\rm AC}=0$, we have $J_0(0)=1$ and $J_k(0)=0$ for $k>0$. Turning on the drive with small amplitude $x=eV_{\rm AC}/(\hbar\Omega_{\rm AC})\ll 1$, $J_0(x)$ decreases quadratically in $x$ whereas $J_1(x)$ increases linearly, and $J_k(x)\propto x^k$ are negligible for $k>1$; since by definition the normalized rate $F(E)$ entering the decay rate increases exponentially for $E$ above the maximal off-voltage $V_{\rm off}$, we have $F(eV_{\rm off}+\hbar\Omega_{\rm AC}+\hbar\omega_{10}) \gg F(eV_{\rm off}+\hbar\omega_{10})$, showing that indeed the QCR can now quickly relax the qubit. On the other hand, the normalized rate entering the excitation rate increases exponentially only if $eV_\mathrm{off} + \hbar\Omega_\mathrm{AC} - \hbar\omega_{10} > E_\mathrm{co}$; therefore, for $\Omega_\mathrm{AC} < 2\omega_{10}$ we have a
regime where the relaxation (excitation) rate is large (small), giving low infidelity and fast reset. As $eV_{\rm AC}/(\hbar\Omega_{\rm AC})$ is further increased, contributions from higher orders $k>1$ also become relevant.

As a concrete example, we calculate numerically infidelity and reset time as functions of the ratio $eV_{\rm AC}/(\hbar\Omega_{\rm AC})$
and of the ac drive frequency $\Omega_{\rm AC}$ for the low-temperature case $k_B T_N/\Delta=0.0043$, using the parameters in Table~\ref{tab:parameters_QCR2level}; the results are presented in Fig.~\ref{fig:AC_drive}. Based on the discussion above, we restrict the ac frequency to $\Omega_\mathrm{AC} < 2\omega_{10} =2\pi \times 10$~GHz. At a small but finite value of $eV_{\rm AC}/(\hbar\Omega_{\rm AC})$, for example $eV_{\rm AC}/(\hbar\Omega_{\rm AC})=0.5$, contributions from orders up to $k=2$ are relevant; we find that the infidelity is lower at $\Omega_{\rm AC}/(2\pi)\approx 5\,$GHz than at $10\,$GHz, 
with an exponential decrease in between when reducing the frequency as the condition $eV_{\rm off}+2\hbar\Omega_{\rm AC}-\hbar\omega_{10} > E_\mathrm{co}$ is not satisfied anymore and hence $F(eV_{\rm off}+2\hbar\Omega_{\rm AC}-\hbar\omega_{10})$ does not contribute significantly to $\Gamma_{01}$. 
By further increasing the ratio $eV_{\rm AC}/(\hbar\Omega_{\rm AC})$, higher-order terms with $k>2$ contribute more and more to the quantities of interest, giving rise to the step-like structures evident in panels (a) and (c);
for our parameters, we find a nearly optimal operating point at $\Omega_{\rm AC}/(2\pi)=1.5$~GHz and $eV_{\rm AC}/(\hbar\Omega_{\rm AC})=4.5$, where we achieve a reset infidelity of $7.8\times 10^{-5}$ within $T_{10\%}\approx40$~ns. These results are comparable to those attainable via dc control with equal QCR parameter values. Therefore, the ac drive provides a viable alternative way to operate the QCR.

\section{Quantum dot QCR coupled to a resonator}
\label{sec:dotQCR}

In the QCR design studied thus far, the normal-metal island is so large that the charging energy is small compared to temperature, $E_N \ll k_B T_N$. This smallness slows down the dynamics of the charge but does not affect the QCR ability to, e.g., reset a qubit. In this section,  we consider the opposite regime of large charging energy and study the system in which the normal-metal island is replaced with a quantum dot. 
The charging energy of the quantum dot is much larger than that of the normal-metal island because of its small size; in fact, we assume that both the single-particle level spacing of the dot $\delta\epsilon$ and the charging energy $E_N$ are large compared to the gap $\Delta$. As a result, only three charge states, the empty state  $\ket{0}$ and the spin up $\ket{\uparrow}$ and down $\ket{\downarrow}$ states are relevant.

\begin{figure}[ht!]
\centering
\includegraphics[width=3.3in]{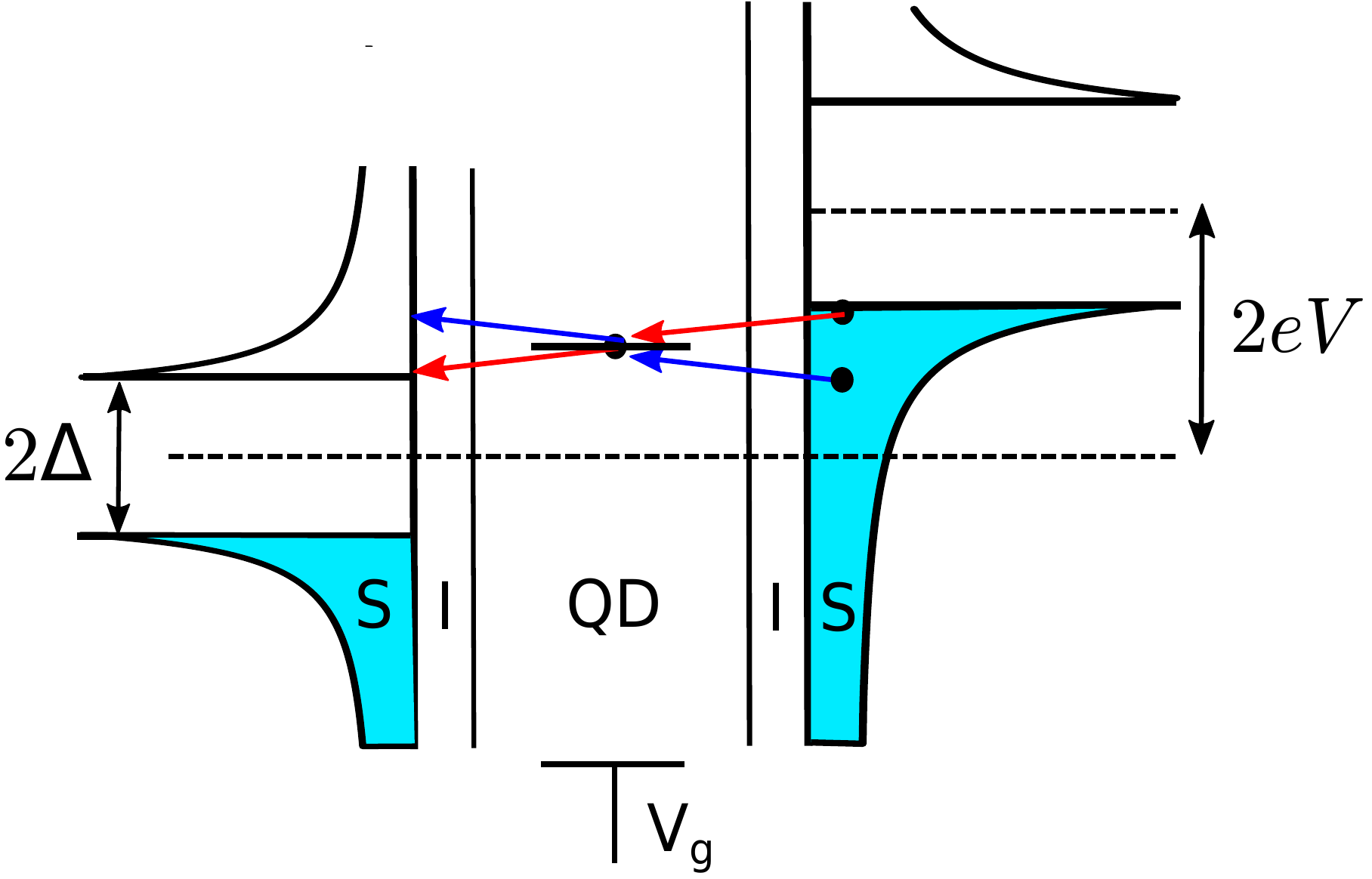} 
\caption{Schematic diagram of the density of states of the voltage biased superconductor--insulator--quantum dot--insulator--superconductor~(SIQDIS) junction. The energy difference between the empty and one-electron states of the quantum dot, which can be tuned by the gate voltage $V_g$, is zero in this figure. Photon-assisted electron tunnelings are depicted by colorful arrows: a blue arrow depicts a tunneling event in which the SIQDIS junction absorbs a photon from the resonator, and a red arrow an event where a photon is emitted into the resonator. Notice that in the depicted biasing regime, the process where a photon is emitted into the resonator (red arrow) is more likely due to the peaked density of states in the superconducting leads. We refer to the bias-voltage-controlled and capacitively coupled SIQDIS junction as the dot QCR.}
\label{fig:dotQCR_resonator}
\end{figure}

The working principle of such a dot QCR is depicted in Fig.~\ref{fig:dotQCR_resonator}, which should be contrasted to Fig.~\ref{fig:QCR_2level}(b). We will consider here the dot QCR connected to a harmonic oscillator. The effective circuit diagram is thus identical to that in Fig.~\ref{fig:QCR_2level}(a) but without the Josephson junction ($E_J=0$). Similarly to the previous section, we will derive the master equation for the diagonal part of the reduced density matrix $\rho_\mu(m)$, where $\mu=0,\, \uparrow,\,\downarrow$ denotes one of the three dot states and $m=0,\,1,\,2,\,\ldots$ the number state in the oscillator.

\subsection{Model}
\label{sec:model_dotQCR}

The Hamiltonian modeling the system can be cast in a form similar to Eq.~(\ref{eq:Htot}),
\begin{equation}
    \hat H_{\rm tot}=\hat H_0+\hat H_\textrm{QD}+\hat H_S+\hat H_T,\\
\end{equation}
with
\begin{align}
     \hat H_0 & =\frac{e^2}{2C_\textrm{QD}}\left(\hat{n} - n_b\right)^2 + \frac{\hat Q_0^2}{2C}+\frac{\hat \Phi_0^2}{2L} \\
     & = E_N \left(\hat{n} - n_b\right)^2 +\hbar \omega_r \left(m+\frac{1}{2}\right) \ket{m}\bra{m} \nonumber
\end{align}
where $C_\textrm{QD}=C_c+C_{\Sigma m}+C_g=C_c+C_m+C_j+C_g$ is the total quantum dot capacitance,  $n_b=C_g V_g/e$ is the dimensionless offset charge, $\hat{n}=\hat{d}^\dagger_{\uparrow}\hat{d}^{\phantom{\dagger}}_\uparrow + \hat{d}^\dagger_{\downarrow}\hat{d}^{\phantom{\dagger}}_\downarrow$ is the dot number operator, $\omega_r=1/\sqrt{LC}$ is the resonator frequency, and $\ket{m}$ are resonator number states. 
The dot charging energy being included in $\hat{H}_0$, the quantum dot Hamiltonian $\hat H_\textrm{QD}$ is simply 
\begin{equation}
    \hat H_\textrm{QD} = \epsilon_\textrm{QD} \sum_{\sigma=\uparrow,\downarrow} \hat{d}^\dagger_{\sigma} \hat{d}^{\phantom{\dagger}}_{\sigma}
\end{equation}
where the dot level energy $\epsilon_\textrm{QD}$ is measured with respect to the equilibrium chemical potential. Since in the following we consider only symmetric bias for the two junctions with respect to the dot level, we may set $\epsilon_\textrm{QD}=0$.
The tunneling Hamiltonian $\hat H_T$ reads
\begin{equation}
    \hat H_T = \sum_{k\sigma} T_k \hat{d}^\dagger_{\sigma} \hat{c}_{k\sigma} e^{i\frac{e}{\hbar}Vt}e^{-i\frac{e}{\hbar}\alpha\phi_0} + \mathrm{H.c.},
    \label{eq:H_T_dotQCR}
\end{equation}
with $\alpha=C_c/C_\textrm{QD}$, and the superconducting leads Hamiltonian $\hat H_S$ is as in Sec.~\ref{sec:QCR}. We assume that the dot and the leads are weakly coupled, so that the normal-state elastic tunneling rate $\gamma_L$
is small compared to the gap, $\gamma_L \ll \Delta$, and therefore Andreev processes can be neglected.\cite{LevyYeyati1997}$^{,}$\footnote{More precisely, our $\gamma_L$ is calculated at zero temperature and finite bias $eV<\delta\epsilon$; our definition differ from that in Ref.~\onlinecite{LevyYeyati1997} by a factor of 2.}
The energy difference between the empty and the one-electron charge states is expressed as $\tilde E_\textrm{QD}=E_N(1-2n_b)$. When the offset charge is $n_b=0.5$, the three charge states have equal energies; this is the charge degeneracy point at which Coulomb blockade is lifted in the normal state.\cite{houten1992}

Using the approach outlined in Sec.~\ref{sec:QCR_master}, we find the master equation (also known as Pauli master equation or rate equations) for the occupation probability $\rho_\mu(m)$ of the resonator state $m$ conditioned on the dot state $\mu$:
\begin{align}
    \dot\rho_\sigma(m) &=  \frac{\gamma_L}{\delta\epsilon}\! \sum\limits_{m',\tau=\pm}\!\! M^2_{mm'}\bigg[ \rho_0(m') F_d(\tau eV-\tilde E_\textrm{QD}-\hbar\omega_r l)\notag\\&-\rho_\sigma(m) F_d(\tau eV+\tilde E_\textrm{QD}+\hbar\omega_r l)\bigg], \label{eq:rho1&3m_EOM} \\
    \dot\rho_0(m) &=  \frac{\gamma_L}{\delta\epsilon}\! \sum\limits_{m',\tau=\pm}\!\! M^2_{mm'}\bigg[ 
   \rho_\uparrow(m') F_d(\tau eV+\tilde E_\textrm{QD}-\hbar\omega_r l)\notag\\
   &+\rho_\downarrow(m') F_d(\tau eV+\tilde E_\textrm{QD}-\hbar\omega_r l)\notag \\
   &-2\rho_0(m) F_d(\tau eV- \tilde E_\textrm{QD}+\hbar\omega_r l)\bigg], \label{eq:rho2m_EOM}
\end{align}
where $\sigma=\uparrow,\,\downarrow$ and  $l=m-m'$. We assume that the superconductor--insulator--quantum-dot and the quantum-dot--insulator--superconductor junctions are identical. The prefactor $\gamma_L/\delta\epsilon$ with the level spacing $\delta\epsilon$ in the denominator is defined as to coincide with $R_K/2R_T$ [cf. Eqs.~(\ref{eq:rho0})--(\ref{eq:rate_qmmp})] in the regime of level spacing being the smallest energy scale. The normalized rate of single-electron tunneling is
\begin{equation}\label{eq:F_dotQCR}
    F_d(E)=\frac{1}{h}n_S(E)\left[ 1-f(E)\right]\delta\epsilon,
\end{equation}
where we assume the quasiparticle excitations of the superconductor to be at thermal equilibrium with temperature $T_S$; as we will see, the relevant energies are of the order of the gap, so that for usual operating temperatures $T_S \ll \Delta$, we can neglect the distribution function term, since $f(E) \ll 1$. Note that because the quantum dots has well-separated energy levels, the normalized rate in Eq.~\eqref{eq:F_dotQCR} probes directly the superconducting density of states and thus displays a sharp peak at $E\simeq \Delta (1+\gamma_D/\sqrt{3})$, in contrast with the smooth behavior of $F(E)$ in Eq.~(\ref{eq:FE_def}).
The matrix elements for the harmonic oscillator can be given explicitly in terms of Laguerre polynomials\cite{Silveri_2017} as functions of the dimensionless impedance normalized by the coupling strength,\footnote{We use here the notation $\zeta$ for the quantity $\rho$ of Ref.~\onlinecite{Silveri_2017} to avoid confusion with the density matrix.}
\begin{equation}\label{eq:z_def}
    \zeta = \pi \alpha^2 \frac{Z_r}{R_K}\, , \quad  Z_r = \sqrt{\frac{L}{C}} \, .
\end{equation}
For later use, we give here only approximate equations for $M^2_{m-1,m}$ valid at $\zeta \ll 1$:
\begin{equation}\label{eq:M1ph}
    M^2_{m-1,m} \approx \left\{ 
    \begin{array}{ll}
       m \zeta (1-m\zeta),  & m\zeta \ll 1  \\
       \frac{\sin^2\left(2\sqrt{m\zeta}-\frac{\pi}{4}+\frac{9}{48\sqrt{m\zeta}}\right)}{\pi\sqrt{m\zeta}},  &  m\zeta \gg 1
    \end{array}
    \right.
\end{equation}
Note that in the regime of interest the matrix elements are approximately functions of the product $m\zeta$, but in general they depend on both $m$ and $\zeta$.

In practical devices, and/or for measurement purposes, the resonator is usually coupled to the environment via a transmission line. This coupling introduces an additional loss mechanism, which we model by adding the following terms to the right sides of Eqs.~\eqref{eq:rho1&3m_EOM}--\eqref{eq:rho2m_EOM} 
\begin{equation}\label{eq:tr_line}
\gamma_{\rm tr}\left[ (m+1)\rho_\mu(m+1)-m\rho_\mu(m)\right],
\end{equation}
where $\mu = \uparrow,\,\downarrow,\, 0$,  and $\gamma_{\rm tr}$ characterizes the strength of the coupling. The latter can be related to the transmission line impedance, resonator impedance, and coupling capacitor -- see Ref.~\onlinecite{Silveri_2017} for details. For simplicity, we assumed a low-temperature environment compared with the resonator frequency, so that the thermal excitations of the resonator can be neglected. 
We show next that the dot QCR can be used to more quickly empty the resonator from photons in comparison to simply waiting for them to leak into the transmission line. This operation regime is the identical to that for refrigeration and qubit reset.\cite{hsu_2020,Silveri_2017} In Sec.~\ref{sec:sol_rate_dotQCR_inv}, a different regime will be discussed in which the dot QCR can be used to pump photons into the resonator; interestingly, non-classical states can be generated in the resonator when pumping by the QCR is balanced by the loss due to the transmission line.

Before proceeding, we note that because of the indistinguishability between spin up and down states in our system (which could be lifted by magnetic field or ferromagnetic junctions, for example), we can simplify the master equation by considering the combinations $\rho_{\pm }(m)=\rho_\uparrow(m)\pm\rho_\downarrow(m)$. The equation for $\rho_-$,
\begin{align}
    \dot\rho_-(m) &= - \frac{\gamma_L}{\delta\epsilon}\! \sum\limits_{m',\tau=\pm}\!\!\! M^2_{mm'}\rho_-(m) F_d(\tau eV+\tilde E_\textrm{QD}+\hbar\omega_r l) \nonumber \\
    & + \gamma_{\rm tr}\left[ (m+1)\rho_-(m+1)-m\rho_-(m)\right]
\end{align}
decouples from those for $\rho_+$ and $\rho_0$ and implies the decay of any spin imbalance; we will not consider this equation any further. The equation for $\rho_+$ is obtained from Eq.~(\ref{eq:rho1&3m_EOM}) with Eq.~(\ref{eq:tr_line}) added to the right side and with the replacementes $\sigma \to +$ and $\rho_0 \to 2\rho_0$.

\subsection{Regime of positive damping}\label{sec:sol_rate_dotQCR_cool}

To estimate the rate at which the number of photons in the resonator decreases when the dot QCR is active, we can restrict our attention to the simplest situation in which there is initially only one photon. Since the goal is to make the decay much faster, in our calculation we can neglect the effect of the transmission line.
Restricting the possible resonator states to $m=0,\, 1$, we write the master equation
in a matrix form
\begin{equation}\label{eq:EOM_4x4}
    \dot{\vec{\rho}}=\boldsymbol{M}\vec{\rho},
\end{equation}
where 
\begin{equation}\label{eq:Matrix_4x4}
   \vec{\rho}= \left( {\begin{array}{*{20}{c}} 
   \rho_+(0)  \\
   \rho_+(1)  \\
   \rho_0(0)  \\
   \rho_0(1)  \\
   \end{array}} \right), \;\;
   \boldsymbol{M} =  \left(\begin{array}{rr}
   -A^+ & 2B^- \\
   B^+ & -2A^-
   \end{array}\right)
\end{equation}
with
\begin{equation}
   A^\pm =  \left(\begin{array}{cc}
   \Gamma^\pm_{00}+\Gamma^\pm_{01} & 0 \\
   0 & \Gamma^\pm_{11}+\Gamma^\pm_{10}
   \end{array}\right) , \quad 
   B^\pm =  \left(\begin{array}{cc}
   \Gamma^\pm_{00} & \Gamma^\pm_{10} \\
   \Gamma^\pm_{01} & \Gamma^\pm_{11}
   \end{array}\right)
\end{equation}
and the transition rates are defined by
\begin{equation}
    \Gamma^{\pm}_{mm'}=\frac{\gamma_L}{\delta\epsilon}M^2_{mm'}\sum\limits_{\tau=\pm}F_d(\tau eV\pm \tilde E_\textrm{QD}+\hbar\omega_r l).
    \label{eq:QD_Gamma_pm_mmp}
\end{equation}

Now let us first consider the charge-degeneracy point where $\tilde E_\textrm{QD}=0$.
For cooling, we want the excitation rates $\Gamma^{\pm}_{01}$ to be much smaller than the decay rates; this can be achieved by requiring $eV + \hbar\omega_r > \Delta$, so that the decay rate is large due to the finite superconducting density of states, and at the same time $eV - \hbar\omega_r < \Delta$, so that the excitation rate is small due to the smallness of the subgap density of states; this defines a voltage window for cooling $\Delta-\hbar\omega_r < eV < \Delta + \hbar\omega_r$. We can therefore neglect the excitation rates. Assuming low resonator impedance $Z_r = \sqrt{L/C} \ll R_K$ 
(which implies $\zeta\ll1$ and hence $M^2_{00}\approx M^2_{11}$,\cite{Silveri_2017}) the dynamical matrix depends only on two parameters
\begin{equation}
    \tilde{\boldsymbol{M}}=
   \left( {\begin{array}{*{20}{c}}
    -\Gamma^0_{00}   &  0    & 2\Gamma^0_{00}  & 2\Gamma^0_{10}  \\
     0   & -\Gamma^0_{00}-\Gamma^0_{10}  & 0   & 2\Gamma^0_{00}  \\
     \Gamma^0_{00}   &  \Gamma^0_{10}    & -2\Gamma^0_{00} & 0   \\
     0   &  \Gamma^0_{00}    & 0   & -2\Gamma^0_{00}-2\Gamma^0_{10}
    \end{array}} \right),
\end{equation}
where $\Gamma^0_{mm'}$ are transition rates for $\tilde E_\textrm{QD}=0$. The eigenvalues of $\tilde{\boldsymbol{M}}$ are $0$, $-3\Gamma^0_{00}$, and
\begin{equation}\label{eq:4x4_appro_Epm}
    E_{\pm}=-\frac{3}{2}\Gamma^0_{00}-\frac{3}{2}\Gamma^0_{10}\pm \frac{1}{2}\sqrt{9(\Gamma^0_{00})^2+2\Gamma^0_{00}\Gamma^0_{10}+(\Gamma^0_{10})^2}.
\end{equation}
The eigenstates corresponding to the two eigenvalues independent of $\Gamma_{10}^0$, $(2/3,0,1/3,0)^T$ and $(1/2,0,-1/2,0)^T$, represent respectively the steady state with no photons in the resonator and equal occupation probability of the three degenerate dot states and the fast decay of charge disequilibrium within the no-photon subspace. When $\Gamma_{10}^0 \ll \Gamma_{00}^0$, the other two eigenvalues are approximately $E_+ \simeq -4\Gamma_{10}^0/3$ and $E_- \simeq -3\Gamma_{00}^0$; the latter rate gives again the fast decay of charge disequilibrium, albeit in the one-photon subspace. The $E_+$ rate is the decay of the one-photon state to no photons, and can be obtained by averaging over  the occupation probabilities of the dot charge states (probabilities 2/3 and 1/3 for states + and 0) the decay rates from 1 to 0 photons given the charge state ($\Gamma_{10}^0$ and $2\Gamma_{10}^0$, respectively). Similarly to the case of a large normal island, we find that for fast elastic transitions, we can effectively treat the charge states as a bath, even though at most one electron can occupy the dot, as the charge steady state is reached quickly, independent of the photon number.
In contrast,
when $\Gamma^0_{00} \to 0$, the eigenvalues $E_{\pm}$, $-\Gamma^0_{10}$ and $-2\Gamma^0_{10}$, coincide with the decay rates of the one electron and the empty states, respectively, showing that the decay rate of the one-photon state depends on the initial state of the dot.

\begin{figure*}[tb]
  \centering
    \includegraphics[width=1\linewidth]{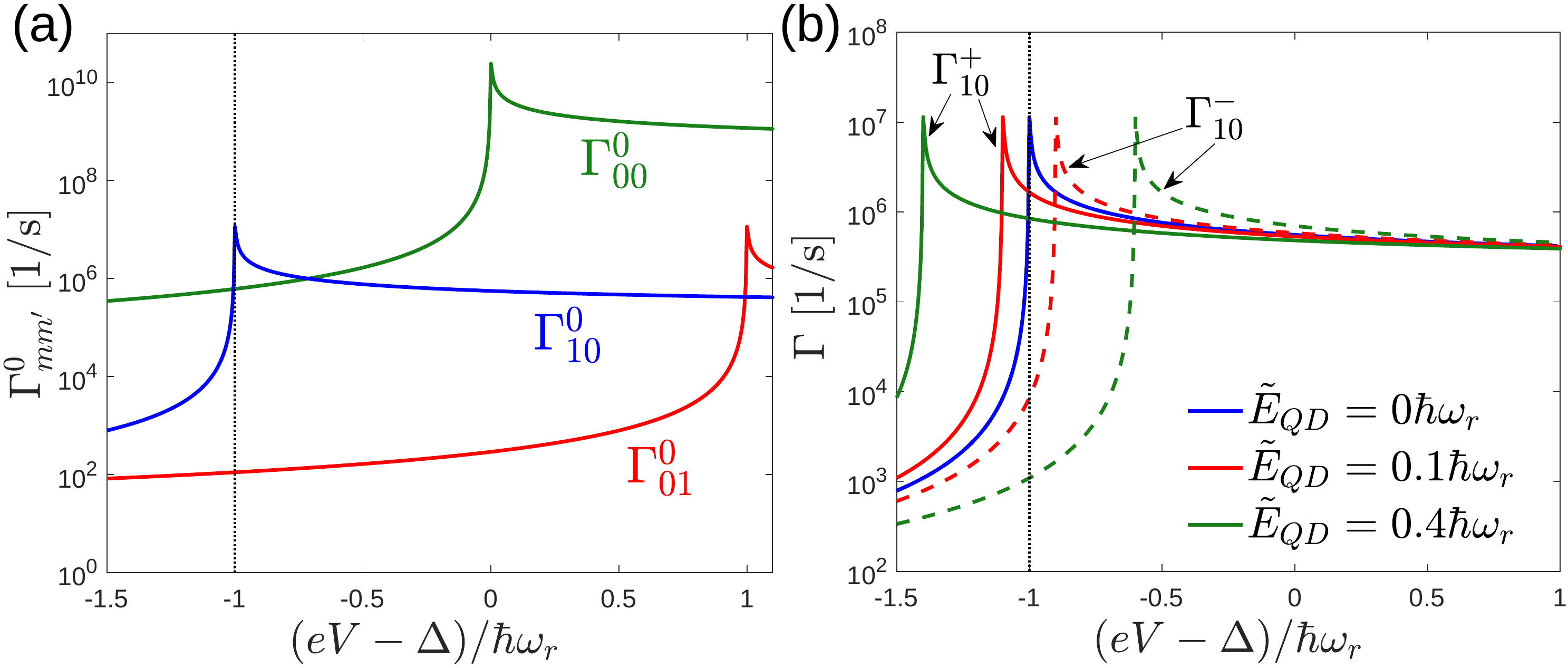}
    \caption{\label{fig:Gamma_pm_mmp_eV}
(a) Transition rates $\Gamma^0_{mm'}$ of Eq.~\eqref{eq:QD_Gamma_pm_mmp} as a function of the bias voltage of the tunnel junction $V$ at charge degeneracy, $\tilde E_\textrm{QD}=0$. (b) Decay rates $\Gamma^{\pm}_{10}$ of Eq.~\eqref{eq:QD_Gamma_pm_mmp} as a function of $eV$ for different $\tilde E_\textrm{QD}$. The solid and dashed lines are for $\Gamma^+_{10}$ and $\Gamma^-_{10}$, respectively. In both (a) and (b), the capacitance ratio is $\alpha = C_c/C_\textrm{QD}=1/3$, and other parameters are given in Table~\ref{tab:parameters_dotQCR}. Note that since the charging energy is large compared to the gap, while the resonator frequency is smaller than the gap, the chosen values of $\tilde{E}_\textrm{QD}$ correspond to a small offset charge deviation from charge degeneracy, $|n_b -1/2| \ll 1$.}
\end{figure*}

We show in Fig.~\ref{fig:Gamma_pm_mmp_eV}(a) the elastic $\Gamma_{00}^0$, decay $\Gamma_{10}^0$, and excitation $\Gamma_{01}^0$ rates as function of bias in the relevant cooling window for realistic device parameters, see Table~\ref{tab:parameters_dotQCR}. Interestingly, due to the peak in the superconducting density of states, there is a region of voltage around $eV \sim \Delta - \hbar\omega_r$ in which the decay rate is larger than the elastic one, while at higher and lower voltages the opposite holds. Note that even at the peak, the predicted relaxation rate is smaller than that estimated in experiment with a metallic-island QCR.\cite{tan_2017,Silveri_2017} Moreover, 
at a fixed bias voltage $V$ the transition rates can strongly depend on the deviation from charge degeneracy, that is, on the value of $\tilde{E}_\textrm{QD} \neq 0$. This is the case for the relaxation rates near $eV \approx \Delta -\hbar\omega_r$, see Fig.~\ref{fig:Gamma_pm_mmp_eV}(b) (cf. also the end of the next subsection). The relaxation rates are weakly dependent on $\tilde{E}_\textrm{QD}$ at higher bias $eV \gtrsim \Delta$, but this decrease in sensitivity to $\tilde{E}_\textrm{QD}$, and hence to charge noise, comes at the cost of a further reduction in the attainable relaxation rate.
We therefore conclude that use of the dot QCR for cooling is not advantageous. We focus next on a regime not possible with the normal-island QCR.

\renewcommand{\arraystretch}{1.25}
\begin{table}[bt]
    \centering
    \begin{tabular}{c c c c c c}
         \multicolumn{1}{c|}{} &   $\delta\epsilon$ & $\gamma_L$ & $\gamma_D$ & $\Delta$ 
         % & $T_S$ 
         \\
         \cline{1-5}
         \multicolumn{1}{c|}{dot QCR} & \SI{1}{\milli\electronvolt} & \SI{2}{\micro\electronvolt} & \num{e-4} & \SI{200}{\micro\electronvolt} 
        %  & \SI{100}{\milli\kelvin} 
         \\
         \\[-0.2cm]
        \multicolumn{1}{c|}{} & $\omega_r/(2\pi)$ & $Z_r$  \\
         \cline{1-3}
        \multicolumn{1}{c|}{Resonator} & \SI{4.67}{\giga\hertz} & \SI{35}{\ohm} \\
         \\[-0.2cm]
    \end{tabular}
    \caption{Typical parameter values for a dot QCR coupled to a resonator. The quantum dot parameters are obtained from Ref.~\onlinecite{Denis_2016}, where a large charging energy ($> 50\,$meV) is reported. The resonator parameters are as in Ref.~\onlinecite{Silveri_2019}.} 
    \label{tab:parameters_dotQCR}
\end{table}

\subsection{Regime of negative damping: analytical approach}
\label{sec:sol_rate_dotQCR_inv}

In the previous subsection we have defined the bias voltage range for cooling $|eV -\Delta| < \hbar\omega_r$ at charge degeneracy $\tilde{E}_\textrm{QD} = 0$. As the voltage increases towards the upper limit of this range, the excitation rate starts to increase [cf. Fig.~\ref{fig:Gamma_pm_mmp_eV}(a)] and in fact, because of the non-monotonicity of the normalized rate $F_d$ of Eq.~(\ref{eq:F_dotQCR}), around the peak at 
\begin{equation}\label{eq:Vp}
    eV_p \simeq \Delta+\hbar\omega_r +\gamma_D \Delta/\sqrt{3}
\end{equation}
the excitation rate is, for typical parameters, much larger than the decay rate, $\Gamma_{01}^0/\Gamma_{10}^0 \approx (3^{3/4}/4\sqrt{\gamma_D}) \sqrt{4\hbar\omega_r/\Delta} \gg 1$. This means that the dot QCR can be used to pump photons into the resonator at a rate faster than that at which it absorbs photons from the resonator. Such pumping cannot be performed with the normal-island QCR, since its normalized rate $F$ [Eq.~(\ref{eq:FE_def})] is a monotonic function: the peak in the superconducting density of states is counteracted by the integration over the flat normal-metal density of states, see Figs.~\ref{fig:QCR_2level} and \ref{fig:dotQCR_resonator}.

For a quantitative understanding of the pumping regime, we return to the master equation (\ref{eq:rho1&3m_EOM}) and (\ref{eq:rho2m_EOM}) which, focusing on the case $\tilde{E}_\textrm{QD} = 0$, we rewrite in terms of $\rho(m)=\rho_+(m)+\rho_0(m)$ and $\rho_z(m)=\rho_+(m)-2\rho_0(m)$; the former represents the probability of having $m$ photons in the resonator, as the dot states are summed over. We now need to retain the loss into the transmission line since, as discussed above, in the pumping regime the QCR-induced excitation rate is larger than the decay rate. The equations are then
\begin{align}
    &\dot\rho(m)=\frac{\gamma_L}{\delta\epsilon} \sum\limits_{m',\tau=\pm1} M^2_{mm'}\bigg[\frac{4}{3}\rho(m')F_d(\tau eV-\hbar\omega_r l)\notag \\
    &-\frac{4}{3}\rho(m)F_d(\tau eV+\hbar\omega_r l)-\frac{1}{3}\rho_z(m') F_d(\tau eV-\hbar\omega_r l) \notag \\&+\frac{1}{3}\rho_z(m) F_d(\tau eV+\hbar\omega_r l) \bigg] \notag \\&+\gamma_{\rm tr}\left[ (m+1)\rho(m+1)-m\rho(m)\right], \label{eq:rhom_EOM} 
    %\\
\end{align}
\begin{align}
    &\dot{\rho_z}(m)=\frac{\gamma_L}{\delta\epsilon} \sum\limits_{m',\tau=\pm1} M^2_{mm'}\bigg[
    -\frac{4}{3}\rho_z(m')F_d(\tau eV-\hbar\omega_r l)\notag \\
    &-\frac{5}{3}\rho_z(m)F_d(\tau eV+\hbar\omega_r l)-\frac{2}{3}\rho(m') F_d(\tau eV-\hbar\omega_r l)  \notag \\
    &+\frac{2}{3}\rho(m) F_d(\tau eV+\hbar\omega_r l)
     \bigg]\notag\\&+\gamma_{\rm tr}\left[ (m+1)\rho_z(m+1)-m\rho_z(m)\right]. \label{eq:rhozm_EOM}
\end{align}
In looking for an approximate solution, we remind that near the voltage bias point $V_p$ of Eq.~(\ref{eq:Vp}) we can neglect the one-photon decay rate in comparison to the one-photon excitation rate. Higher photon decay processes are negligible for the same reason, while higher photon excitation transition are further suppressed by the small sub-gap density of states in the superconducting leads. As for the elastic rates, they cannot in general be neglected, because at low photon number their matrix element $M_{mm}^2 \approx 1 - (2m+1)\zeta$ (valid for $m\zeta \ll 1$) is much larger than the one-photon matrix element, see Eq.~(\ref{eq:M1ph}). Keeping only one-photon excitation and elastic transitions, as well as loss into the transmission line, in the steady state $\dot\rho=\dot\rho_z=0$, the master equation simplifies to
\begin{align}
    0 & = \frac{4}{3}M^2_{-1}\rho(m-1)-\frac{4}{3}M^2_{1}\rho(m)\nonumber \\ & -\frac{1}{3}M^2_{-1}\rho_z(m-1)+\frac{1}{3}M^2_{1}\rho_z(m) \nonumber \\
    & +\tilde\gamma_{\rm tr}\left[ (m+1)\rho(m+1)-m\rho(m)\right] \label{eq:rho_simp} \\
    0 & = -3d_0 M^2_{0}\rho_z(m)-\frac{4}{3}M^2_{-1}\rho_z(m-1) \nonumber \\ & -\frac{5}{3}M^2_{1}\rho_z(m) -\frac{2}{3}M_{-1}^2\rho (m-1)\nonumber \\ & +\frac{2}{3}M_{1}^2\rho (m)\nonumber \\ & +\tilde\gamma_{\rm tr}\left[ (m+1)\rho_z(m+1)-m\rho_z(m)\right] \label{eq:rhoz_simp}
\end{align}
where we used the short-hand notation $M^2_i = M^2_{m,m+i}$ and introduced the dimensionless parameters
\begin{equation}\label{eq:tildegtr}
    \tilde{\gamma}_\mathrm{tr} = \gamma_\mathrm{tr} \left[\frac{\gamma_L}{\delta\epsilon}\sum_{\tau=\pm1} F_d\left(\tau eV_p - \hbar\omega_r\right)\right]^{-1} \approx \frac{4\gamma_\mathrm{tr}\sqrt{\gamma_D}}{3^{3/4}\gamma_L} \,
\end{equation}
and
\begin{equation}\label{eq:d0_def}
    d_0 = \frac{\sum_{\tau=\pm1} F_d\left(\tau eV_p\right)}{ 
\sum_{\tau=\pm1} F_d\left(\tau eV_p - \hbar\omega_r\right)} \approx \frac{4}{3^{3/4}} \sqrt{\frac{\gamma_D\Delta}{2\hbar\omega_r}} \, .
\end{equation}
For typical parameters, $d_0\ll 1$ since the resonator frequency is high compared with the broadening of the density of states, $\omega_r \gg \gamma_D \Delta/\hbar$.
It should be noted that since in simplifying the master equation we have assumed the one-photon transitions to be dominant, the simplification is not applicable for photon numbers near those values at which the oscillatory nature of the matrix element $M^2_{m,m-1}$ [cf. Eq~(\ref{eq:M1ph})] renders it sufficiently small (the simplification is valid at small $m$ though); as we will see below, we are not concerned with those values.

Thanks to its structure, Eq.~(\ref{eq:rho_simp}) assumes a more compact form when approximating it as a differential equation: by summing it to the equation obtained by shifting $m\to m-1$ and neglecting derivatives in $m$ of third order or higher, we find
\begin{equation}\label{eq:rhod_2}
    0 \approx \partial_m \!\!\left[\frac{4}{3}M^2_{-1}\rho(m-1) - \frac{1}{3}M^2_{-1}\rho_z(m-1) - \tilde{\gamma}_\mathrm{tr} m \rho(m) \right]
\end{equation}
Furthermore, treating Eq.~(\ref{eq:rhoz_simp}) as an equation for $\rho_z$ given $\rho$, with the same procedure and accuracy~\footnote{Equation~(\ref{eq:rhozd_sol}) shows that if first or higher order derivatives acting on terms containing $\rho_z$ were kept, by iterative solution third or higher order derivatives acting on terms containing $\rho$ would be introduced in Eq.~(\ref{eq:rhod_2}), terms which would be beyond the used accuracy.} we obtain
\begin{equation}\label{eq:rhozd_sol}
    \left(M^2_{-1} + d_0 M^2_{0}\right) \rho_z(m-1) \approx \frac{2}{9} \partial_m \left[M^2_{-1}\rho(m-1)\right]
\end{equation}
The term proportional to $d_0$ is in most cases negligible: for $m \zeta \gg 1$ and $m$ values where our simplification applies, the matrix elements $M^2_{m,m-1}$ and $M^2_{mm}$ are of equal order of magnitude while $d_0$ is small; only if $m\zeta \lesssim \sqrt{\gamma_D}$ is the $d_0$ term relevant, a situation possible only if $\sqrt{\gamma_D} \gtrsim \zeta$. We drop this term for now and comment on its role later in this section.

Neglecting the $d_0$ term, substituting Eq.~(\ref{eq:rhozd_sol}) into Eq.~(\ref{eq:rhod_2}) we obtain
\begin{equation}\label{eq:rho_las_fin}
    0 = \frac{4}{3}M^2_{-1}\rho(m-1) - \frac{2}{27}\partial_m \left[M^2_{-1}\rho(m-1)\right] - \tilde{\gamma}_\mathrm{tr} m \rho(m)
\end{equation}
where the integration constant has been set to zero since $\rho(m)$ must decay sufficiently fast at large $m$. We now focus on the regime of small photon number, $m\zeta \ll 1$. Then using Eq.~(\ref{eq:M1ph}) we approximate Eq.~(\ref{eq:rho_las_fin}) as
\begin{align}
    0 & = \left[\frac{4}{3}\zeta\left(1-\zeta m\right) - \tilde{\gamma}_\mathrm{tr} \right]\rho(m-1) \\ & - \left[\frac{2}{27}\zeta\left(1-\zeta m\right) + \tilde{\gamma}_\mathrm{tr}\right]
    \partial_m \rho(m-1) \nonumber
\end{align}
The solution to this equation is a Gaussian, $\rho(m) \propto e^{-(m-\bar{m})^2/(2\sigma^2_m)}$, centered at the mean photon number
\begin{equation}\label{eq:m0_low}
    \bar{m} = \frac{1}{\zeta}\left(1-\frac{3\tilde{\gamma}_\mathrm{tr}}{4\zeta}\right)
\end{equation}
and with variance
\begin{equation}\label{eq:sd_lowm}
    \sigma^2_m = \frac{19}{18} \bar{m} \frac{ 3\tilde\gamma_\mathrm{tr}/4\zeta}{1-3\tilde\gamma_\mathrm{tr}/4\zeta}
\end{equation}
These expressions are valid when $|1-3\tilde\gamma_\mathrm{tr}/4\zeta| \ll 1$, and they show that if $3\tilde\gamma_\mathrm{tr}/4\zeta > 1$, then $\bar{m} < 0$, meaning that the probability distribution has a maximum at $m=0$ because the relatively strong coupling to the transmission line effectively counteracts the dot QCR pumping. For weaker coupling to the transmission line,
\begin{equation}\label{eq:gtr1}
    \gamma_\mathrm{tr} < \gamma_\mathrm{tr}^{(1)} \equiv \frac{\gamma_L\zeta}{3^{1/4}\sqrt{\gamma_D}} =\frac{\pi\gamma_L}{3^{1/4}\sqrt{\gamma_D}}\left(\frac{C_c}{C_\textrm{QD}}\right)^2 \frac{Z_r}{R_K}
\end{equation}
the probability maximum is at $\bar{m}>0$ and $\bar{m}$ increases with decreasing $\tilde\gamma_\mathrm{tr}$, indicating that pumping becomes more and more effective as the coupling to the transmission line is decreased. It can be shown that the above condition for effective pumping is unchanged even if, at small $\bar{m}$, the term proportional to $d_0$ in Eq.~(\ref{eq:rhozd_sol}) is taken into consideration. 

\begin{figure}[tb]
  \centering
    \includegraphics[width=1\linewidth]{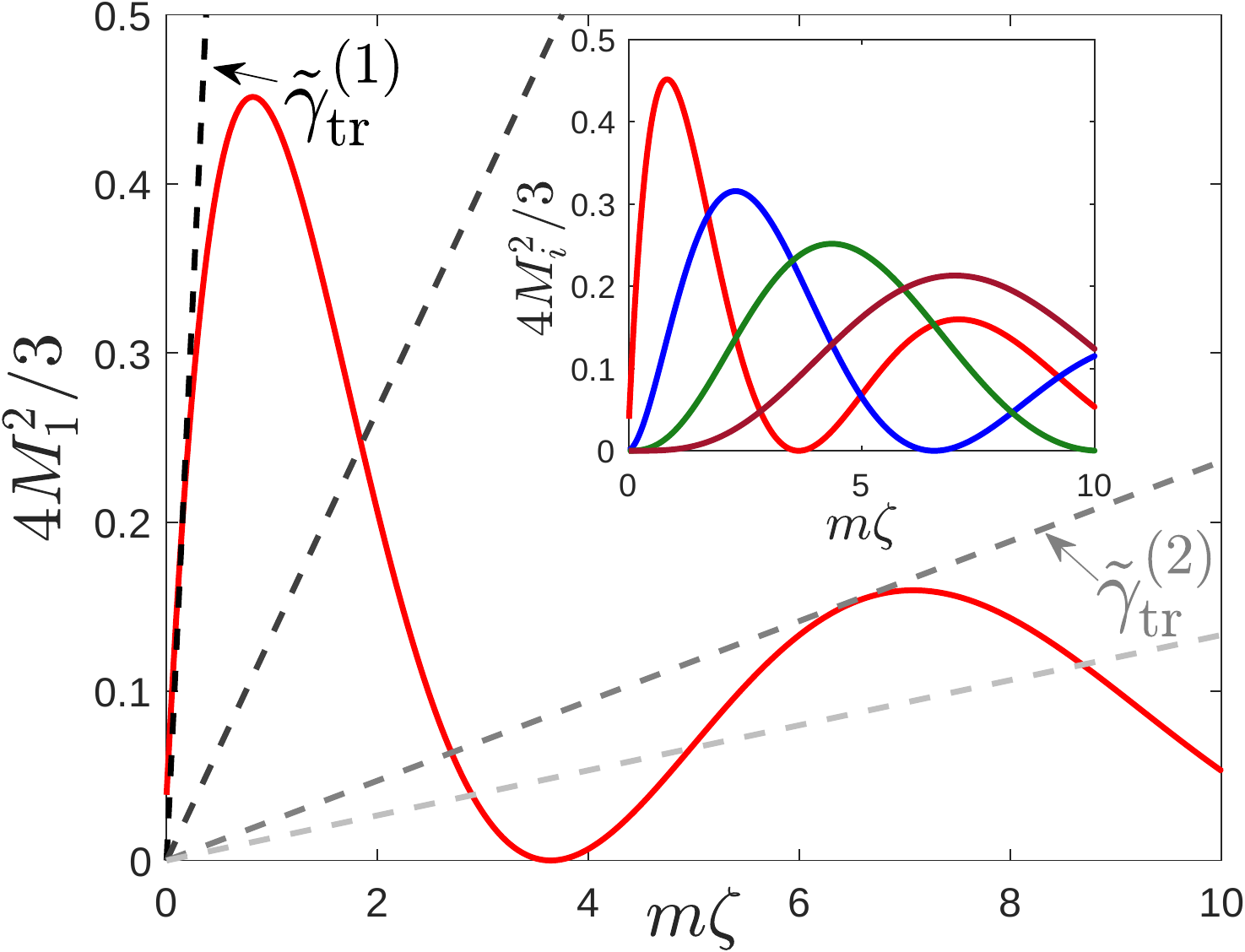}
    \caption{\label{fig:M2_i_mzeta}Main panel: matrix element $4M^2_1/3$ (solid line). Dashed lines are straight lines $\tilde{\gamma}_\mathrm{tr} m$ with different slopes; $\tilde{\gamma}^{(1)}_\mathrm{tr}$ and $\tilde{\gamma}^{(2)}_\mathrm{tr}$ are related to $\gamma^{(1)}_{\rm tr}$ of Eq.~\eqref{eq:gtr1} and $\gamma^{(2)}_{\rm tr}$ of Eq.~\eqref{eq:gtr2} via Eq.~\eqref{eq:tildegtr}. Note that the dashed line with the smallest slope intersects the solid curve multiple times. Inset: matrix elements $4M^2_1/3$, $4M^2_2/3$, $4M^2_3/3$, and $4M^2_4/3$ in order from steeper to flatter near the origin.}
\end{figure}

A useful quantity to consider is the Fano factor 
\begin{equation}
    \mathrm{F} = \frac{\sigma_m^2}{\bar{m}} \, ,
\end{equation}
since $F<1$ signals that the resonator hosts a non-classical state of light with sub-Poissonian statistics.\cite{Fox} In its regime of validity, $m\zeta \ll 1$, Eq.~(\ref{eq:sd_lowm}) predicts that the Fano factor
is large, $\mathrm{F} \gg 1$, but decreases with decreasing $\tilde\gamma_\mathrm{tr}$. We can extend the above considerations to large $m$ ($m\zeta \gg 1$) as follows: by multiplying and dividing the last term in Eq.~(\ref{eq:rho_las_fin}) by $M^2_1$, we can reinterpret that equation as a first-order differential equation for $g(m) = M^2_{-1}\rho(m-1)$:
\begin{equation}\label{eq:gm}
 0 = \left(\frac{4}{3}-\frac{\tilde{\gamma}_\mathrm{tr}m}{M^2_1}\right) g(m) -\left(\frac{2}{27}+\frac{\tilde{\gamma}_\mathrm{tr}m}{M_1^2}\right)\partial_m g(m)
\end{equation}
The coefficient in round brackets multiplying the first-order derivative is always positive. In contrast, the coefficient in front of $g$ can change sign, and when the weak-coupling condition (\ref{eq:gtr1}) is satisfied 
there exists $\bar{m}>0$ solving the equation
\begin{equation}\label{eq:m0g}
    \frac{4}{3}M^2_{\bar{m},\bar{m}+1} - \tilde{\gamma}_\mathrm{tr} \bar{m} = 0
\end{equation}
The term proportional to $\tilde\gamma_\mathrm{tr}$ characterizes the transmission line-induced loss and the matrix element term the gain due to the photon-assisted tunneling; their equality defines the mean photon number where the gain and loss are balanced. As the coupling to the transmission line gets weaker, there can be multiple solutions to this equation, see Fig.~\ref{fig:M2_i_mzeta}. Here we focus only on the regime of single solution~\footnote{The numerical coefficient relating $\gamma_\mathrm{tr}^{(2)}$ to $\gamma_\mathrm{tr}^{(1)}$ is approximately given by $[1-\sin(4y+9/24y)]/(2\pi y^3)$, where $y\approx 2.57$ is the second non-zero, positive solution to $1-\sin(4y+9/24y)=-4y\cos(4y+9/24y)/3$.},
\begin{equation}\label{eq:gtr2}
   \gamma_\mathrm{tr}^{(1)}  > \gamma_\mathrm{tr} > \gamma_\mathrm{tr}^{(2)} \approx 0.017 \gamma_\mathrm{tr}^{(1)}
\end{equation}
which spans almost two orders of magnitude in the coupling strength to the transmission line.

For $m$ near $\bar{m}$, Eq.~(\ref{eq:gm}) can be approximated as
\begin{equation}\label{eq:gm_app}
    0 = -a (m-\bar{m}) g(m) - \partial_m g(m)
\end{equation}
where
\begin{equation}\label{eq:a_def}
a = \frac{18}{19 \bar{m}}\left(1-\bar{m} \frac{\partial_m M^2_1}{M^2_1}\bigg|_{\bar{m}}\right)
\end{equation}
When the conditions in Eq.~(\ref{eq:gtr2}) are met, $a$ is positive~\footnote{For even weaker coupling to the transmission line, the number of solutions to Eq.~(\ref{eq:m0g}) is generally odd and $a$ alternates its sign from one solution to the next, suggesting possible multi-stability regimes. A multi-peak distribution for phonon occupation has been predicted in an optomechanical system, but at low phonon number,\cite{ash_2012} and in a spin-based quantum dot-resonator system,\cite{Gianluca_multi_2019} when treated beyond the secular approximation (which we employ). We do not explore this possibility further in this work.}, and the solution to Eq.~(\ref{eq:gm_app}) is a Gaussian centered at $\bar{m}$ with variance $\sigma_m^2 = 1/a$. The Fano factor
\begin{align}
    \label{eq:Fanog}
    \mathrm{F} & = \frac{1}{a\bar{m}} \\ & \approx \frac{19}{27}\left[1+\frac{4}{3}\frac{\sqrt{\bar{m}\zeta}\cos\left(4\sqrt{\bar{m}\zeta}+9/24\sqrt{\bar{m}\zeta}\right)}{1-\sin\left(4\sqrt{\bar{m}\zeta}+9/24\sqrt{\bar{m}\zeta}\right)}\right]^{-1} \nonumber
\end{align}
is smaller than unity in the validity regime ($\bar{m}\zeta \gtrsim 1$), and becomes as small as $\mathrm{F}\approx 0.14$ as $\gamma_\mathrm{tr} \to \gamma_\mathrm{tr}^{(2)}$~\footnote{In this limit, the value of $\bar{m}\zeta = x^2 \approx 2.72$ to be used in calculating $\mathrm{F}$ is obtained as the second largest positive solution to $1-\sin(4x+9/24x) = 2\pi x^3 \gamma_\mathrm{tr}^{(2)}/\gamma_\mathrm{tr}^{(1)}$}. Note that the cross-over from super-Poissonian ($\mathrm{F}>1$) to sub-Poissonian ($\mathrm{F}<1$) value for the the variance takes place around $\bar{m} \zeta \sim 1$, corresponding to the value where $M_1^2$ is maximal and hence its derivative changes sign, as one could expect from Eq.~(\ref{eq:a_def}). The non-monotonicity of $M_1^2$ also implies a counterintuitive non-monotonic dependence of $\bar{m}$ on the coupling strength between dot QCR and resonator via the parameter $\zeta$: at small $\zeta$, the condition (\ref{eq:gtr1}) is not met, $\bar{m} < 0$, and the distribution is maximal at $m=0$; as $\zeta$ increases, Eq.~(\ref{eq:gtr1}) is satisfied, $\bar{m}$ becomes positive and increases rapidly with $\zeta$ [cf. Eq.~(\ref{eq:m0_low})]; at $\zeta=\zeta_\mathrm{max} \simeq 1.875 \tilde\gamma_\mathrm{tr}$, $\bar{m}$ reaches its maximum value $\bar{m}_{\mathrm{max}}\approx 0.85/\zeta_\mathrm{max}$~\footnote{The numerical coefficients in these relations have been determined numerically by locating the first maximum of the matrix element $M^2_1$ as function of $m\zeta$.}; finally, as $\zeta$ is further increased, $\bar{m}$ slowly decreases.

The above analysis indicates that the dot QCR can induce non-classical states of light in the resonator, with $\mathrm{F}<1$, which may be useful in metrology because of their narrow photon distribution.\cite{tan_2019} Above, we have neglected the possibility of photon absorption by the QCR; in the next section we study numerically its effect as well as that of deviation from charge degeneracy. Here we note that the main effect of single-photon absorption by the dot QCR is to suppress the matrix element in Eq.~(\ref{eq:m0g}) by a factor $1-d_1$ with
\begin{equation}\label{eq:d1_def}
    d_1 = \frac{\sum_{\tau=\pm1} F_d\left(\tau eV_p + \hbar\omega_r\right)}{ 
\sum_{\tau=\pm1} F_d\left(\tau eV_p - \hbar\omega_r\right)} \approx \frac{d_0}{\sqrt{2}}
\end{equation}
and $d_0$ of Eq.~(\ref{eq:d0_def}).
This suppression leads to a small decrease in $\bar{m}$ and hence to a small increase of the Fano factor. The absorption of $n$ photons has a similar effect, but in the regime $\bar{m} \lesssim 1/\zeta$ is quantitatively weaker due to the suppression of the matrix element by $\sim \zeta^{n-1}$ in comparison to the single photon transitions, cf. the inset in Fig.~\ref{fig:M2_i_mzeta}. This figure also shows that for larger $\bar{m}$, the two- and three-photon absorption processes must be taken into account, whereas we can neglect transitions with $n\ge 4$ provided that we stay in the single-solution regime defined by Eq.~(\ref{eq:gtr2}). The results of this section remain approximately valid for small deviations from charge degeneracy, $|\tilde{E}_\textrm{QD}| \lesssim \gamma_D \Delta$, upon the substitution $\tilde\gamma_\mathrm{tr} \to \tilde\gamma_\mathrm{tr}^{(s)}$, where
\begin{equation}\label{eq:gtr_renorm}
    \tilde{\gamma}_\mathrm{tr}^{(s)} = \gamma_\mathrm{tr} \left[\frac{\gamma_L}{2\delta\epsilon}\sum_{\tau,\upsilon=\pm1} F_d\left(\tau eV_p +\upsilon \tilde{E}_\textrm{QD} - \hbar\omega_r\right)\right]^{-1},
\end{equation}
a result that can be obtained by showing that, in the $\tilde{E}_\textrm{QD}$ range mentioned above, the effect of the antisymmetric (in $\tilde{E}_\textrm{QD}$) part of $F_d$ in Eqs.~(\ref{eq:rho1&3m_EOM}) and (\ref{eq:rho2m_EOM}) can be neglected.
Since the term in square brackets is smaller than the corresponding one in Eq.~(\ref{eq:tildegtr}), the average photon number $\bar{m}$ decreases, and the Fano factor $\mathrm{F}$ increases, with the deviation from charge degeneracy.

\subsection{Regime of negative damping: numerical results and microwave generation}
\label{sec:pumping2}

\begin{figure*}[bt]
  \centering
    \includegraphics[width=0.95\linewidth]{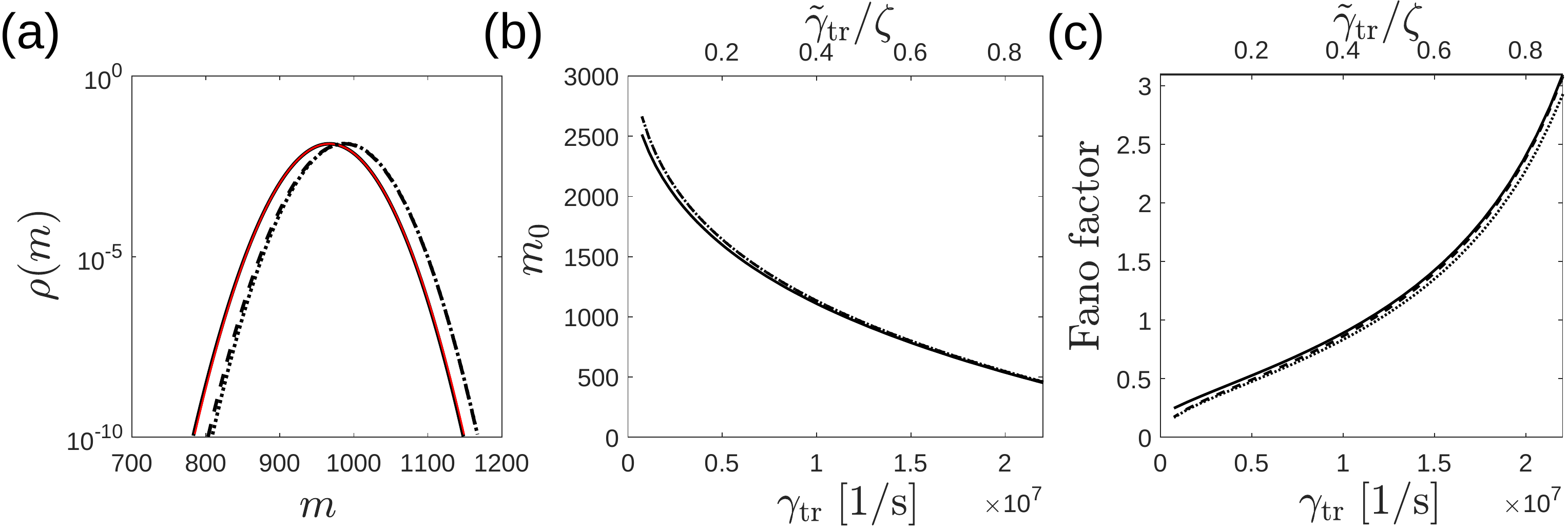}
    \caption{\label{fig:rhom_m0_Fano} (a) Photon distribution $\rho(m)$ as a function of $m$. The black solid (dashed) curve is the numerical result including up to three (single-)photon transitions, and the red solid curve is its Gaussian fitting. The dotted curve is the analytical Gaussian result, where the mean photon number and the variance are obtained using Eqs.~\eqref{eq:m0g} and \eqref{eq:a_def}, respectively, with the correction factor in Eq.~\eqref{eq:d1_def} included. (b) Mean photon number $\bar{m}$ and (c) Fano factor $\mathrm{F}$ as functions of the resonator-transmission line coupling strength $\gamma_{\rm tr}$ (bottom axis) or the dimensionless parameter $\tilde\gamma_{\rm tr}/\zeta$ (upper axis), see  Eqs.~\eqref{eq:z_def} and \eqref{eq:tildegtr}. Solid (dashed), and dotted lines are numerical results including up to three- (single-)photon  transitions and analytical results, respectively. The Fano factor is calculated by using Eq.~\eqref{eq:m0g} and Eq.~\eqref{eq:a_def} in Eq.~\eqref{eq:Fanog}. In this figure, we have set $\zeta=0.001$, which corresponds to using $\alpha\simeq 0.48$ in Eq.~\eqref{eq:z_def}. For panel (a), $\gamma_\mathrm{tr} = 1.2\times 10^7\,$s$^{-1}$; the other parameters are given in Table~\ref{tab:parameters_dotQCR}.}
\end{figure*}

Having explored analytically the solution to the master equation \eqref{eq:rho1&3m_EOM}--\eqref{eq:rho2m_EOM} with the inclusion of the transmission line  [cf. Eqs.~\eqref{eq:rhom_EOM}--\eqref{eq:rhozm_EOM}] in the pumping regime near charge degeneracy, we now turn to its numerical solution for quantitative predictions. Focusing first on the charge degenerate case, we compare our analytical results (dotted curves) to the numerical ones (dashed) obtained by retaining only elastic and one-photon transitions. In Fig.~\ref{fig:rhom_m0_Fano}(a), we show a typical example for the photon number distribution $\rho(m)$; only small deviations from the analytical (Gaussian) prediction are visible, validating our approach. In Figs.~\ref{fig:rhom_m0_Fano}(b) and (c), we show the average photon number $\bar{m}$ and the Fano factor $\mathrm{F}$, respectively, as functions of the coupling strength $\gamma_\mathrm{tr}$ to the transmission line. As the coupling increases, $\bar{m}$ decreases and $\mathrm{F}$ increases, and good agreement is evident between numerical and analytical results.

We can include in our numerical calculations the effects of 2- and 3-photon transitions; the results are shown as solid lines in Fig.~\ref{fig:rhom_m0_Fano}. In the three panels, we find that the average photon number slightly decreases and the Fano factor slightly increases compared to the one-photon case, as qualitatively predicted in the previous subsection. In panels (b) and (c), the relative differences between dotted and solid lines are larger at small $\gamma_\mathrm{tr}$; this is due to the larger relevance there of multi-photon transitions, see the matrix elements in the inset of Fig.~\ref{fig:M2_i_mzeta}.

\begin{figure}[bt]
  \centering
    \includegraphics[width=0.9\linewidth]{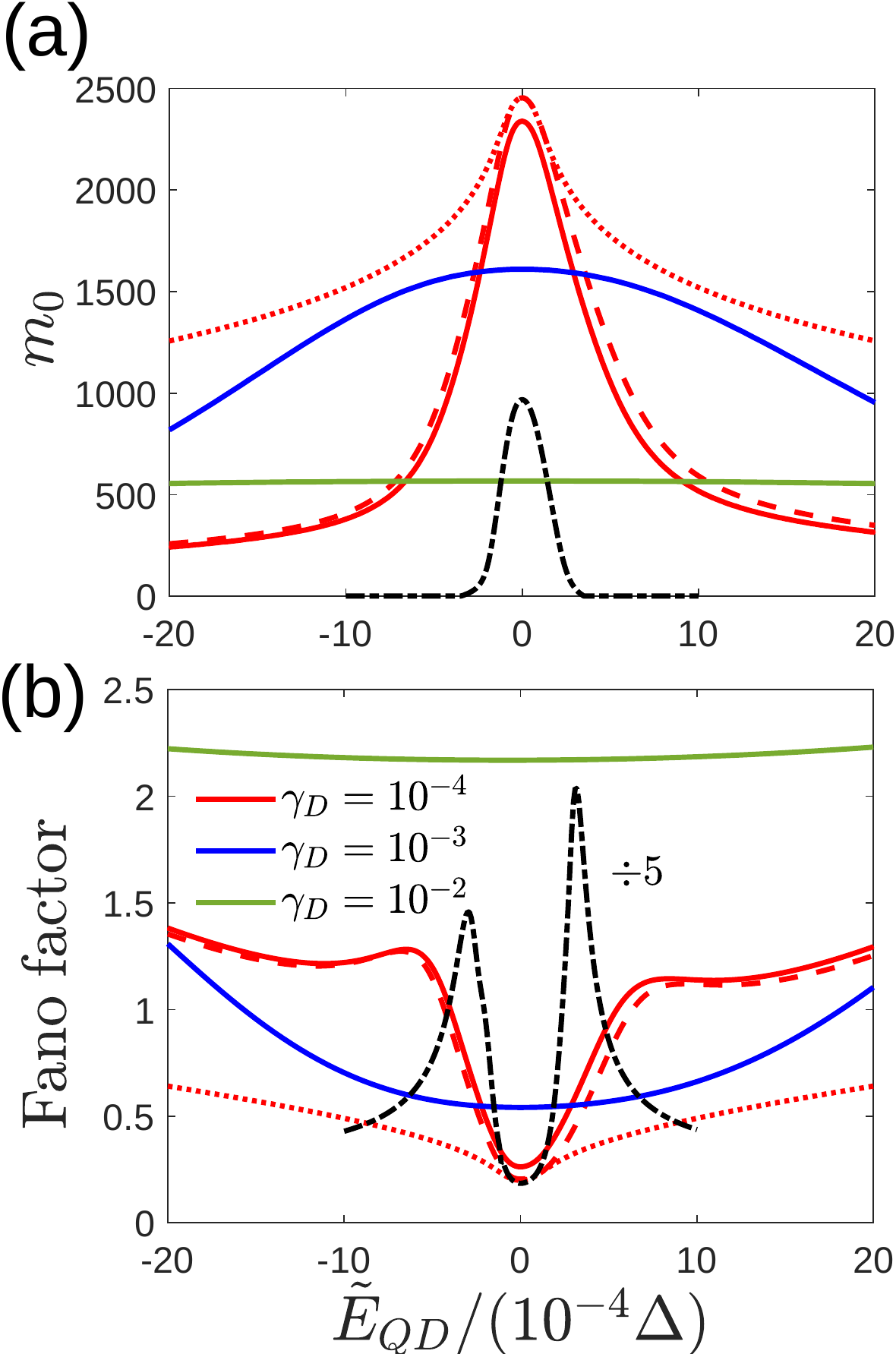}
    \caption{\label{fig:m0_Fano_EQD} Average photon number $\bar{m}$ (a) and Fano factor $\mathrm{F}$ (b) as functions of the deviation from charge degeneracy $\tilde{E}_\textrm{QD} = E_N(1-2n_b)$. Solid lines: numerical results setting $\zeta=0.001$ (as in Fig.~\ref{fig:rhom_m0_Fano}), $\gamma_\mathrm{tr} = 1.2\times 10^6\,$s$^{-1}$, and the other parameters (except for $\gamma_D$) as given in Table~\ref{tab:parameters_dotQCR}. Dashed lines: numerical results with equal parameter values but including only elastic and one photon transitions. Dotted lines: analytical results based on the replacement in Eq.~(\ref{eq:gtr_renorm}). Dot-dashed lines: numerics with $\gamma_D = 10^{-4}$ and $\gamma_\mathrm{tr} = 1.2\times 10^7\,$s$^{-1}$; in panel (b) the results have been suppressed by a factor of 5 for clarity.}
\end{figure}

We can also investigate the deviation from charge degeneracy, $\tilde{E}_\textrm{QD} \neq 0$. As expected, $\bar{m}$ decreases and $\mathrm{F}$ increases as $|\tilde{E}_\textrm{QD}|$ increases, see Fig.~\ref{fig:m0_Fano_EQD}. Within the approximations made (retaining only up to single photon transitions and assuming $|\tilde{E}_\textrm{QD}| \lesssim \gamma_D \Delta$), the replacement in Eq.~(\ref{eq:gtr_renorm}) leads to results that agree with the numerical ones (dot and dashed lines, respectively). However, with the inclusion of higher photon processes and at larger deviations from charge degeneracy the effects are more pronounced. The sensitivity on the value of $\tilde{E}_\textrm{QD}$ can be suppressed by raising $\gamma_D$, at the price of further reduction of $\bar{m}$ and increase in $\mathrm{F}$ at charge degeneracy. We  return to this point below after discussing negative damping. In agreement with the above analysis at charge degeneracy, strengthening the coupling to the transmission line leads to overall decrease in $\bar{m}$ and increase in $\mathrm{F}$, see the dot-dashed curves in Fig.~\ref{fig:m0_Fano_EQD}.

Above, we focused on the state of the cavity. However, as previously remarked, the steady-state is reached by balancing the dot QCR pumping with the loss into the transmission line; in other words, radiation is emitted by the cavity.  The rate at which the photons leak into the transmission line is $\gamma_\textrm{tr}\sum_m m \rho(m) \approx \gamma_\textrm{tr} \bar{m}$. Multiplication of this rate by the photon energy yields the generated power
\begin{equation}\label{eq:power}
    P= \gamma_\textrm{tr} \bar{m}\hbar\omega_r.
\end{equation}
Based on the analysis in the previous section and our numerical results, we conclude that the maximum power is reached at charge degeneracy and limited by the (first) maximum value $M_{1,\mathrm{max}}^2 \approx 0.34$ of the matrix element $M^2_1$ to be
\begin{equation}
    P_\mathrm{max} \simeq 3^{-1/4} M_{1,\mathrm{max}}^2 \frac{\gamma_L}{\sqrt{\gamma_D}} \hbar\omega_r
\end{equation}
Unsurprisingly, $P_\mathrm{max}$ grows with the transition rate between superconducting leads and quantum dot; it also grows as the superconducting density of states becomes more peaked. For an order-of-magnitude estimate we use the parameters in Table~\ref{tab:parameters_dotQCR} to get $P_\mathrm{max} \sim38\,$~fW. This maximum power can be delivered only if the dot QCR-cavity coupling and cavity-transmission line coupling are appropriately matched, $\zeta/\gamma_\mathrm{tr} \simeq 3.29 \sqrt{\gamma_D}/\gamma_L$~\footnote{The parameters used in Fig.~\ref{fig:rhom_m0_Fano}(a) are close to satisfying this condition, which can be met by increasing $\gamma_\mathrm{tr}$ to $1.47\times10^7\,$Hz.}. In this case, the Fano factor is of order unity; therefore, depending on the application it may be advantageous to weaken the coupling to the transmission line $\gamma_\mathrm{tr}$ to below the matching value: while this decreases the power, it would also brings the Fano factor below unity, as previously predicted for single-atom,\cite{mu_1992} superconductor/quantum dot/normal lead,\cite{rastelli_2019} and transmon-based\cite{astafiev_2021} microwave sources. Interestingly, if $\gamma_\mathrm{tr}$ is increased instead, the dynamics of the system is dominated by single-photon transitions only, and the behaviors of mean photon number and Fano factor as functions of deviation from charge degeneracy (dot-dashed lines in Fig.~\ref{fig:m0_Fano_EQD}) resemble those calculated for a biased double quantum dot coupled to a resonator as functions of detuning between resonator frequency and the frequency of the (effective) two-level system describing the dots.\cite{jin_2011,xu_2013}

A potential issue in using the dot QCR-cavity system as a microwave source is its possible sensitivity to charge noise, similar to that observed in masers based on a superconducting qubit\cite{astafiev_2007} and on double quantum dots.\cite{liu_2015} Indeed, in the previous paragraph we assumed charge degeneracy, but $\bar{m}$ (and hence the power) quickly decreases, while the Fano factor increases, as the charge degeneracy condition $n_b=1/2$ is violated; in fact, under our assumptions we require $1-2n_b$ to be small compared to $\gamma_D$. Increasing $\gamma_D$ by two orders of magnitude decreases the sensitivity to charge noise by an equal amount, while decreasing the maximum power by only one order of magnitude; in principle, the Dynes parameter can be increased by using normal/superconductor bilayers in place of the superconducting electrodes.\cite{Hosseinkhani} Alternatively, the use of a superconductor with higher gap than Al would also decrease the charge noise sensitivity while only slightly decreasing the maximum power, as the latter is largely determined by the coupling to transmission line. Taking Nb or NbN as examples, this would likely result in junctions with larger $\gamma_D\sim 10^{-2}$ as well.\cite{Nevala,Chaudhuri}

\begin{figure}[bt]
  \centering
    \includegraphics[width=0.9\linewidth]{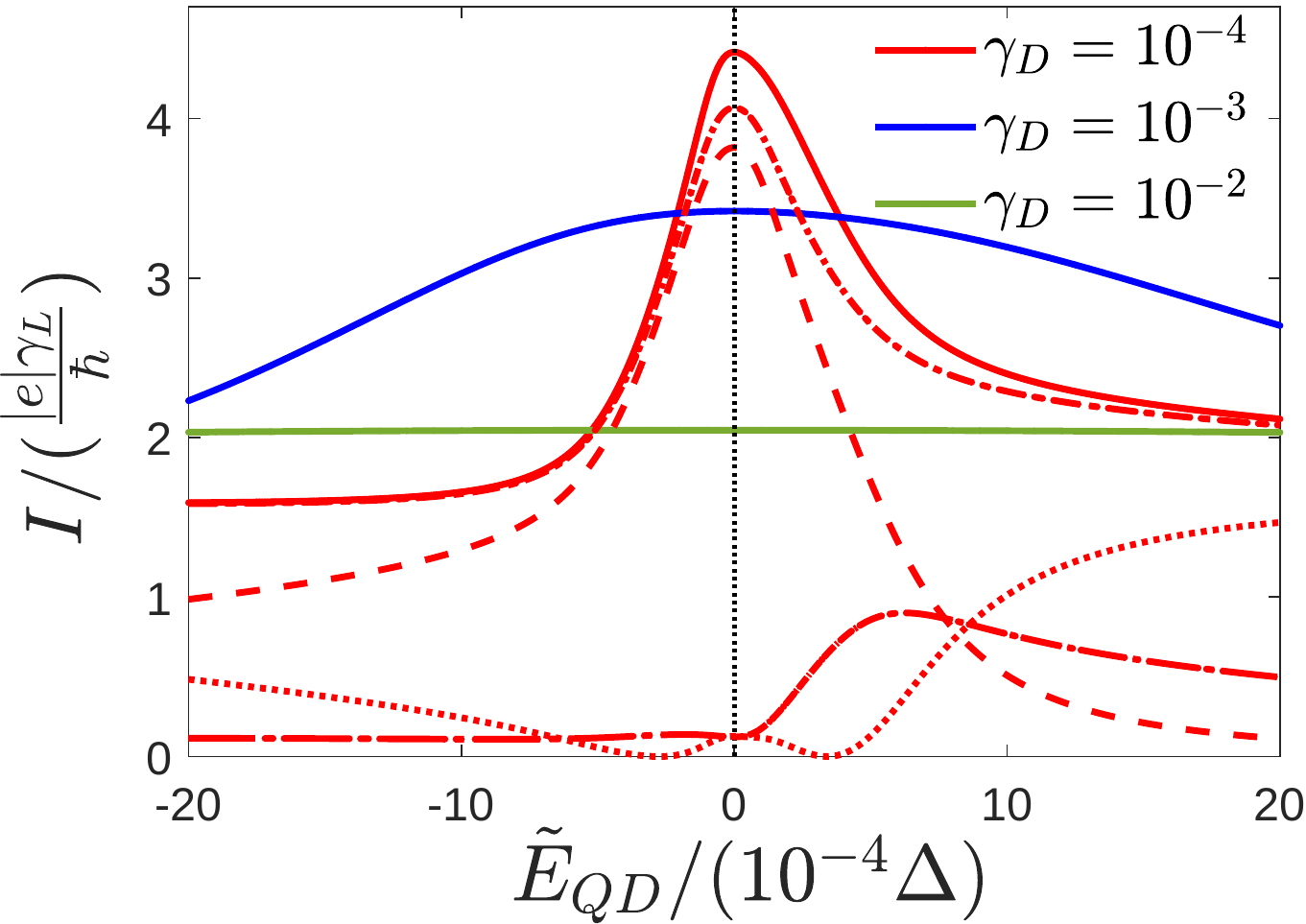}
    \caption{\label{fig:current_EQD} Normalized current $I$ as a function of the deviation from charge degeneracy $\tilde{E}_\textrm{QD} = E_N(1-2n_b)$. Solid lines: numerical results of the total current using equal parameter values to those in Fig.~\ref{fig:m0_Fano_EQD}(a). Dotted, dashed, dotted long-dashed, and dotted short-dashed lines are also calculated with these parameters (but only for $\gamma_D=10^{-4}$) and represent the following contributions to the current originating respectively from: elastic transitions, inelastic 1-photon emission, 1-photon absorption, and the sum of these three terms.
    }
\end{figure}

As we show, it may be possible to stabilize the source against charge noise by monitoring the current through the quantum dot and then adjusting the gate voltage accordingly. The current $I_R$ leaving the right lead ($I_L$ entering the left lead) can be calculated as the time derivative of the electron number in that lead:\cite{Barone}
\begin{align}
I_R  = & \frac{2\pi|e|\gamma_L}{\delta\varepsilon} \sum_{m,m'} M^2_{mm'} \bigg[ 2\rho_0(m') F_d(eV-\tilde E_\textrm{QD}-\hbar\omega_r l)\notag\\&-\rho_+(m) F_d(- eV+\tilde E_\textrm{QD}+\hbar\omega_r l)\bigg] \\
I_L = & \frac{2\pi|e|\gamma_L}{\delta\varepsilon} \sum_{m,m'} M^2_{mm'} \bigg[ \rho_+(m) F_d(eV+\tilde E_\textrm{QD}+\hbar\omega_r l)\notag\\&-2\rho_0(m') F_d(- eV-\tilde E_\textrm{QD}-\hbar\omega_r l)\bigg]
\end{align}
with $l=m-m'$. In the steady state, it follows from the master equation [cf. Eqs.~(\ref{eq:rho1&3m_EOM}) and (\ref{eq:rho2m_EOM})] that the two currents are equal, $I_R=I_L \equiv I$. In Fig.~\ref{fig:current_EQD}, we show with solid lines the numerical results for the normalized current as function of $\tilde{E}_\textrm{QD}$, using the parameter values as in Fig.~\ref{fig:m0_Fano_EQD}(a). The two plots are similar, with the current maximum being very close in position to $\tilde{E}_\textrm{QD}=0$, a result resembling again that for a double quantum dot system.\cite{xu_2013} In our case, the similarity between the two plots is a consequence of the current's two main contributions originating from the inelastic 1-photon transitions from the QCR (dominant near charge degeneracy) plus the elastic transitions (relevant away from degeneracy), see the dotted/dashed lines in Fig.~\ref{fig:current_EQD} . Therefore, adjusting the gate voltage as to maximize the current through the dot will also keep the output power at its maximum value.

\section{Summary and outlook}
\label{sec:conclusions}

In this work, we study the QCR by deriving a master equation for the reduced system density matrix. This allows us to investigate the dynamics of excess charge on the QCR normal-metal island that is coupled to the qubit. The charge relaxation rates, which characterize how fast the excess charge distribution of the QCR reaches the steady state, are found to be much lower than the qubit reset rates. It validates the approximations carried out in our previous work where we calculated the transition rates by Fermi's golden rule.\cite{Silveri_2017, hsu_2020}
Based on our theory, we also study ac-voltage control of the QCR, with the control-signal frequency comparable to the frequencies of the superconducting qubits. We find that with typical parameter values it is possible to achieve a reset infidelity of $7.8\times 10^{-5}$ with the order-of-magnitude reset time of $T_{10\%}\approx40$~ns. This is comparable to previous results achieved by pure dc control\cite{sevriuk_2019} and provides a simple complementary approach to that in Ref.~\onlinecite{viitanen}.

We also use the master equation approach to study a modified QCR coupled to a resonator, where a quantum dot takes the place of the normal-metal island; the charging energy and the level spacing of the quantum dot are assumed to be larger than the superconducting gap. Due to the corresponding constraints on the capacitance between the QCR and the resonator and the lower number of electron states in on the QCR island, the tunneling rates are significantly lower than in the original metallic QCR. This may prevent the utilization of such device for qubit reset. In addition, our results show that this variant of the QCR may be sensitive to charge noise.

Interestingly, biasing the quantum dot QCR with an appropriate voltage, we find that the excitation rates are much higher than the relaxation rates. Thus, the dot QCR provides effectively negative damping into the microwave resonator, a phenomenon observed in other types of devices such as voltage-biased Josephson junctions\cite{Yan2021} to lead to coherent microwave emission.
In this regime we can solve the master equation under the assumption that the excitation rates are dominant. We also find numerical solutions for the master equation which are consistent with our analytical results.
The maximum generated power achievable with realistic experimental parameters is $\sim 38$~fW, and at the maximum power the Fano factor is of order unity. 
However, non-classical states of light with Fano factor smaller than unity can be obtained, although with lower power. The source is susceptible to charge noise, which is a typical problem of quantum-dot lasers. We propose here that feedback control of the dot gate voltage based on monitoring the tunneling current will help stabilize the output of the source.

Interesting future research directions include the possibility of phase locking of the QCR microwae source in a similar fashion as has been shown in Ref.~\onlinecite{Yan2021} for a Josephson junction source. For microwave technologies and quantum-information research, it may be fruitful to investigate the use of this device for cryogenic signal generation, and at least theoretically establish its basic parameters as an amplifier. In comparison to the Josephson junction source where the noise temperature of the source output is governed by the electron temperature of the shunt resistor, the quantum dot may provide voltage control over this temperature owing to the discrete energy spectrum.

\begin{acknowledgments}
This research was financially supported by the Academy of Finland under Grant Nos. 316619, 319579, 316551, and~335460 and under its Centres of Excellence Programme (Grant No. 312300), by the Internationalization Fund -- Cutting-Edge Ideas initiative of Forschungszentrum
J\"ulich, by the Centre for Quantum Engineering at Aalto under Grant number JVSG, and by the European Research
Council under Grant Nos. 681311 (QUESS) and~680167 (SCAR).
M.M. is a Founder of IQM which is a commercial quantum-computer company.
\end{acknowledgments}

\section*{Data Availability Statement}

The data that support the findings of this study are available from the corresponding author upon reasonable request.

\appendix

\section{Master equation at second order in \texorpdfstring{$E_N/T_N$}{En/Tn}}
\label{app:2nd}

\begin{figure}[tb]
  \centering
  \includegraphics[width=1.0\linewidth]{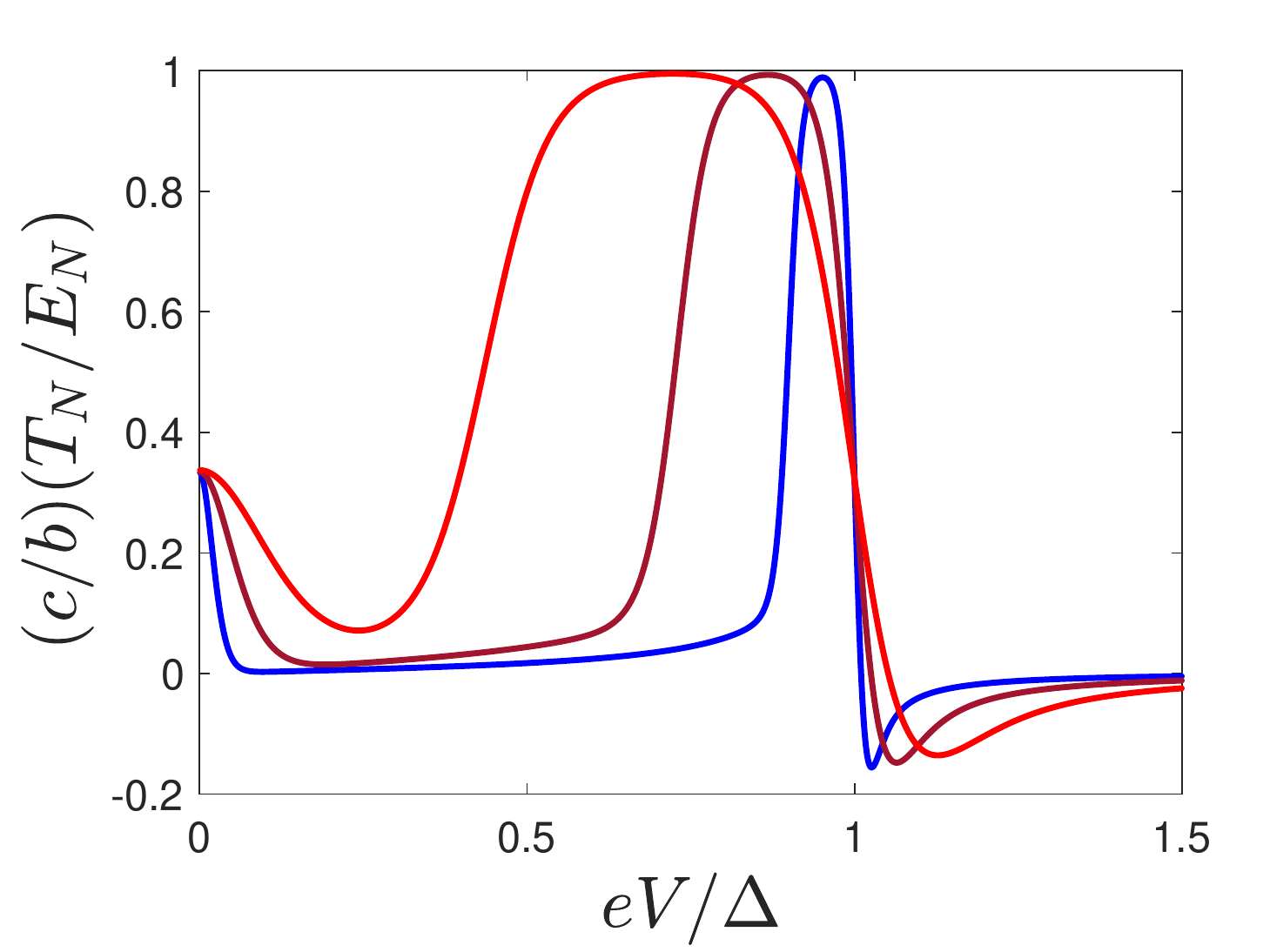}
  \caption{\label{fig:ratio_eV} The ratio of $c$ of Eq.~\eqref{eq:c} to $b$ of Eq.~\eqref{eq:b} multiplied by $T_N/E_N$ as a function of $eV$ for different island temperatures, corresponding to $E_N/T_N \approx 0.009$ (red), $0.005$ (brown), and $0.002$ (blue).}
\end{figure}

In looking for an approximate solution to the master equation, in Sec.~\ref{sec:sol_master_QCR} we consider the first order expansion of the transition rates in the small factor $b \sim E_N/T_N \ll 1$. We argue there that for the inelastic rate, only the zeroth order terms are important, while we keep both zeroth and first order terms for the elastic rates. Here we give further evidence that such an expansion is justified by including the second order terms for the elastic rates, while completely neglecting the inelastic ones for simplicity. More precisely, we assume $M_{00} = M_{11}$ (so that $\eta =0$) and $M_{01}=0$; then in rescaled time units, Eq.~(\ref{eq:Gel}), all three components of the master equation, Eqs.~(\ref{eq:rho0}), (\ref{eq:rho1}), and (\ref{eq:rho+}) [in the rotating frame], take the same form, cf. Eqs.~(\ref{eq:rho0_1st_expandF})--(\ref{eq:rho+_elastic_1st}),
\begin{align}
\dot{\rho}(q)& =\left[1+b(1+2q)+\frac{1}{2}c (1+2q)^2\right]\rho(q+1) \\
&+\left[1+b(1-2q)+\frac{1}{2}c (1-2q)^2\right]\rho(q-1) \nonumber \\
& -2\left[1-b+\frac{1}{2}c \left(1+4q^2\right)\right]\rho(q) \nonumber
\end{align}
where
\begin{equation}
    c=E_N^2 \frac{F''(eV)+F''(-eV)}{F(eV)+F(-eV)}.
    \label{eq:c}
\end{equation}
It can be shown that $|c| < b E_N/T_N$ and therefore $|c| \ll b$, see Fig.~\ref{fig:ratio_eV}

Following the steps that lead to Eq.~(\ref{eq:barrhoj}), with the canonical transformation now being $p=\tilde{p} + 4i\tilde{m}(b+c)$ with $2\tilde{m}=1/(1+b+c/2)$, we arrive at
\begin{align}   \label{eq:rho0_HO_2nd}
    &\lambda_j \rho_{j} =\left[\frac{\tilde p^2}{2\tilde{m}}+\frac{1}{2}\tilde{m}\tilde{\omega}^2 q^2-2b+2c\right]\rho_{j}  \\
    &+\left[2cq^2 \tilde{p}^2+16i\tilde{m}c(b+c)q^3 \tilde{p}-32\tilde{m}^2c(b+c)^2 q^4\right]\rho_j , \nonumber
\end{align}
where $\tilde{\omega}^2 = 16 (b+c)(b+2c)$. The terms in the second line can be recognized as anharmonic corrections to the harmonic oscillator Hamiltonian in the first line. Introducing as usual the creation and annihilation operators via $q= (a^\dagger+a)/\sqrt{2\tilde{m}\tilde{\omega}}$ and $p=i(a^\dagger-a)\sqrt{\tilde{m}\tilde{\omega}/2}$, and considering the second line within first-order perturbation theory, at first order in $c$ we find
\begin{equation}   \label{eq:eigenvalue_2nd}
    \lambda_j = 4bj -2cj(j+1)
\end{equation}
This shows that for slowly decaying modes with $j \ll T_N/E_N$ the corrections to the decay rates due to higher order terms can be neglected.

\section{Corrections to the distribution function}
\label{app:charge_correction}

We state in Sec.~\ref{sec:sol_master_QCR} that the correction terms $\delta\bar\rho_j$ and $\delta\tilde\rho_j$ are small for $\eta,\,\gamma \ll1$. Here we show why this is the case, and also comment on the validity of the approximation introduced using Eq.~(\ref{eq:Taylor}).

Let us introduce the notations
\begin{align}
 \mathcal{D}\left[\rho\right] = & \rho(q+1) + \rho(q-1) - 2\rho(q) \\
    \mathcal{D}_b\left[\rho\right]  = & \left[1+b(1+2q)\right] \rho(q+1) \label{eq:Db_def} \\  +  & \left[1+b(1-2q)\right] \rho(q-1)  - 2(1-b)\rho(q) \,. \nonumber
\end{align}
After substituting the Ans\"atze (\ref{eq:charge_ans}) and (\ref{eq:q_ans}) into Eqs.~(\ref{eq:rho0_1st_expandF}) and (\ref{eq:rho1_1st_expandF}), we obtain the following two sets of coupled equations:
\begin{align}
 -\lambda_j  \bar\rho_j = & \, \mathcal{D}_b\left[\bar\rho_j\right] + \frac{\bar\Gamma}{2} \mathcal{D}\left[\bar\rho_j\right]  \label{eq:rhob_app} \\
& +\frac{\gamma}{4} \mathcal{D}\left[\delta\bar\rho_j\right] + \gamma \delta\bar\rho_j(q) -\eta\mathcal{D}_b\left[\delta\bar\rho_j\right] \nonumber \\
-\lambda_j \delta\bar\rho_j = & \, \mathcal{D}_b\left[\delta\bar\rho_j\right]
-\frac{\bar\Gamma}{2} \mathcal{D}\left[\delta\bar\rho_j \right] -2\bar\Gamma\delta\bar\rho_j(q)  \label{eq:drhob_app} \\
& -\frac{\gamma}{4} \mathcal{D}\left[\bar\rho_j \right] -\eta \mathcal{D}_b\left[\bar\rho_j\right] \nonumber
\end{align}
and 
\begin{align}
-\lambda_j  \tilde\rho_j = & \, \mathcal{D}_b\left[\tilde\rho_j\right]-\frac{\bar\Gamma}{2}\mathcal{D}\left[\tilde\rho_j\right]-2\bar\Gamma \tilde\rho_j(q)  \label{eq:rhot_app} \\
& +\frac{\gamma}{4}\mathcal{D}\left[\delta\tilde\rho_j\right] + \gamma\delta\tilde\rho_j(q) -\eta\mathcal{D}_b \left[\delta\tilde\rho_j\right]\nonumber \\
-\lambda_j  \delta\tilde\rho_j = & \, \mathcal{D}_b\left[\delta\tilde\rho_j\right]
+ \frac{\bar\Gamma}{2}\mathcal{D}\left[\delta\tilde\rho_j\right] \label{eq:drhot_app} \\
& -\frac{\gamma}{4}\mathcal{D}\left[\tilde\rho_j\right]-\eta\mathcal{D}_b \left[\tilde\rho_j\right]
\nonumber  
\end{align}
It is clear that in the limits $\eta,\, \gamma \to 0$ we can set the corrections to zero, $\delta\bar\rho_j = \delta\tilde\rho_j =0$. Therefore for $\eta,\, \gamma \ll 1$ one can solve the equations by iterations: first one solves Eqs.~(\ref{eq:rhob_app}) and (\ref{eq:rhot_app}) for $\bar\rho_j$ and $\tilde\rho_j$ neglecting the corrections terms in their second lines; those solutions provide non-homogeneous terms in Eqs.~(\ref{eq:drhob_app}) and (\ref{eq:drhot_app}) for the corrections. After solving the latter equations, one could then improve the solution by substituting the first-order corrections back into Eqs.~(\ref{eq:rhob_app}) and (\ref{eq:rhot_app}) to find second-order terms, and so on. We do not pursue this further here, and turns instead our attention to the approximation in Eq.~(\ref{eq:Taylor}).

Even setting $\gamma=\eta=0$, Eqs.~(\ref{eq:rhob_app}) and (\ref{eq:rhot_app}) are not identical. A trivial difference is the third term on the right side of Eq.~(\ref{eq:rhot_app}), which leads to the increase by $2\bar\Gamma$ of the decay rates discussed below Eq.~(\ref{eq:rhot}), but does not influence the dependence of the solution on $q$ (in the harmonic oscillator picture, it amounts to a shift in the reference point for energy). On the contrary, the second terms on the right sides of the two equations have opposite signs and affect the charge dependence, but not the decay rates; the latter are calculated in Sec.~\ref{sec:sol_master_QCR} for $\gamma\ll 1$ and arbitrary value of $\bar\Gamma$. Here we assume the more stringent condition $\bar\Gamma \ll 1$, which implies $\gamma \ll 1$; then we can treat the terms proportional to $\bar\Gamma$ as small perturbations, and an iterative solution can be constructed for the average and difference of $\bar\rho_j$ and $\tilde\rho_j$, in a way similar to that described in the previous paragraph.

We focus here on the main equation for the average $\rho_j = (\bar\rho_j+\tilde\rho_j)/2$, which reads simply
\begin{equation}\label{eq:rhoj_app}
    -\lambda_j \rho_j = \mathcal{D}_b \left[\rho_j\right]
\end{equation}
with $\mathcal{D}_b$ of Eq.~(\ref{eq:Db_def}). Since the small parameter $b$ appears there multiplied by $1\pm2q$, one might expect that an approximate solution can be found when $bq \ll 1$, which is a more stringent condition than the necessary one, $b\ll1$. This requirement is confirmed by calculating the ratio between successive terms in the expansion in Eq.~(\ref{eq:Taylor}) for the two slowest approximate solutions, $\rho_0 = e^{-2bq^2}$ and  $\rho_1=q e^{-2bq^2}$ (up to normalization) -- solutions that are straightforwardly guessed from the harmonic oscillator analogy. Improved solutions can be obtained with the replacement $e^{-2b q^2} \to e^{-2b(1-b^2/3)q^2 - 4b^3 q^4/3}$, as can be checked by substitution into Eq.~(\ref{eq:rhoj_app}) and subsequent Taylor expansion in $b$. Clearly, the last term in the exponent gives a small correction when $(bq)^2 \ll 1$, while suppressing the tails of the distribution more strongly than in a Gaussian distribution.  We conclude by noting that numerical implementation and solution of the full equations (\ref{eq:rho0_1st_expandF}) and (\ref{eq:rho1_1st_expandF}) is straightforward, and confirms all the results presented here and in the main text.

\section{Total qubit decoherence}
\label{app:totdec}

The inclusion of non-QCR decoherence mechanisms into the master equation, Eqs.~(\ref{eq:rho0}), (\ref{eq:rho1}), and (\ref{eq:rho+}) is straightforward: the corresponding rates do not depend on the QCR charge and the qubit transitions do not change the charge state. Therefore, we simply add the terms $\Gamma_d^{(0)} \rho_1(q) - \Gamma_u^{(0)} \rho_0(q)$ to the right-hand side of Eq.~(\ref{eq:rho0}), $\Gamma_u^{(0)} \rho_0(q) - \Gamma_d^{(0)} \rho_1(q)$ to the right-hand side of Eq.~(\ref{eq:rho1}), and $-\left(\frac12\Gamma_u^{(0)}+\frac12 \Gamma_d^{(0)} +\Gamma_\varphi^{(0)}
\right)\rho_+(q)$ to the right-hand side of Eq.~(\ref{eq:rho+}). Hereinafter we use superscript $(0)$ to denote non-QCR rates, while as above subscript $u$ denotes excitation, $d$ decay, and $\varphi$ pure dephasing. The same terms [with rates now expressed in units of the elastic rate $\Gamma_{el}$ of Eq.~(\ref{eq:Gel})] are added to the right-hand sides of Eqs.~(\ref{eq:rho0_1st_expandF})--(\ref{eq:rho+_elastic_1st}). For the latter equation, this gives simply the expected increase in each decay rate $\lambda_j$ by the non-QCR mechanisms. We focus next on Eqs.~(\ref{eq:rho0_1st_expandF})--(\ref{eq:rho1_1st_expandF}).

The above additions do not alter the structure of Eqs.~(\ref{eq:rhob_app})--(\ref{eq:drhot_app}), but lead to a redefinition of some of the coefficients appearing there once the Ansatz in Eq.~(\ref{eq:charge_ans}) is modified to 
\begin{align}
    \rho_{0j}(q) & = \frac{\Gamma_d+\Gamma_d^{(0)}}{\Gamma_u+\Gamma_d+\Gamma_u^{(0)}+\Gamma_d^{(0)}}\left[\bar\rho_j(q) + \delta\bar\rho_j(q)\right], \\
    \rho_{1j}(q) & = \frac{\Gamma_u+\Gamma_u^{(0)}}{\Gamma_u+\Gamma_d+\Gamma_u^{(0)}+\Gamma_d^{(0)}}\left[\bar\rho_j(q) - \delta\bar\rho_j(q)\right]. \nonumber
\end{align}
while Eq.~(\ref{eq:q_ans}) is unchanged. Then in Eqs.~(\ref{eq:rhob_app}) and (\ref{eq:drhob_app}) one should make the replacements
\begin{align}
  \frac{\bar\Gamma}{2}\mathcal{D}[\cdot] & \to \frac{\bar\Gamma_1}{2}\mathcal{D}[\cdot] \\
  \frac{\gamma}{4}\mathcal{D}[\cdot] & \to \frac{\gamma_1}{4}\mathcal{D}[\cdot] \\
  \gamma \delta\bar\rho_j & \to \gamma_2 \delta\bar\rho_j \\
  2\bar\Gamma \delta\bar\rho_j & \to 2\bar\Gamma_2 \delta\bar\rho_j
\end{align}
with
\begin{align}
    \bar\Gamma_1 & = \frac12 \left[\frac{\Gamma_u}{\Gamma_u+\Gamma_u^{(0)}}\left(\Gamma_d + \Gamma_d^{(0)}\right) + d\leftrightarrow u\right]\label{eq:Gamma1_app}\\
    \bar\Gamma_2 & =  \frac12 \left[\left(\Gamma_d + \Gamma_d^{(0)}\right)  +  d\leftrightarrow u \right] \\
    \gamma_1 & = \frac{\Gamma_u}{\Gamma_u+\Gamma_u^{(0)}}\left(\Gamma_d + \Gamma_d^{(0)}\right) -  d\leftrightarrow u\\
    \gamma_2 & =\left(\Gamma_d + \Gamma_d^{(0)}\right)  -  d\leftrightarrow u 
    \label{eq:gamma2_app}
\end{align}
In Eq.~(\ref{eq:rhot_app}) the replacements are
\begin{align}
  \gamma \delta\tilde\rho_j & \to \gamma_2 \delta\tilde\rho_j \\
  2\bar\Gamma \tilde\rho_j & \to 2\bar\Gamma_2 \tilde\rho_j
\end{align}
and Eq.~(\ref{eq:drhot_app}) is unaffected. Since the revised equations differ from the original ones only in their parameters, all the considerations Appendix~\ref{app:charge_correction} and in the main text remain unchanged, so long as the dimensionless rates in Eqs.~(\ref{eq:Gamma1_app})--(\ref{eq:gamma2_app}) are in the perturbative regime of being small compared to unity. In particular, we find that the decay rates discussed below Eq.~(\ref{eq:barrhoj}) do not change, while those below Eq.~(\ref{eq:rhot}) increase by the term $\Gamma_u^{0}+\Gamma_d^{0}$, thus proving the claim at the end of Sec.~\ref{sec:sol_master_QCR} that only the temporal evolution of $\rho_2(t)$ in Eq.~(\ref{eq:rhofactor}) is altered by the non-QCR decoherence mechanisms.

%\normalem

\bibliography{references}{}

\end{document}